\documentclass[12pt]{article} 

\usepackage{url,hyperref,lineno,microtype,subcaption}
\usepackage[onehalfspacing]{setspace}

\usepackage{physics,amsmath,amsfonts,amssymb,graphicx,amsthm,geometry}
\usepackage{stackrel,stmaryrd,etoolbox,framed,accents,mathrsfs,upgreek}
\usepackage[all]{xy}

\def\be{\begin{equation}}
\def\ee{\end{equation}}
\def\ba{\begin{eqnarray}}
\def\ea{\end{eqnarray}}

\makeatletter
\newsavebox{\@brx}
\newcommand{\llangle}[1][]{\savebox{\@brx}{\(\m@th{#1\langle}\)}%
  \mathopen{\copy\@brx\kern-0.5\wd\@brx\usebox{\@brx}}}
\newcommand{\rrangle}[1][]{\savebox{\@brx}{\(\m@th{#1\rangle}\)}%
  \mathclose{\copy\@brx\kern-0.5\wd\@brx\usebox{\@brx}}}
\makeatother

\newgeometry{vmargin={25mm}, hmargin={22mm,22mm}} 

\numberwithin{equation}{section}

\begin{document}

\setcounter{tocdepth}{1}

\title{Asymptotic behaviour of massless fields \\and kinematic duality between\\ interior null cones and null infinity} 

\author{Xavier Bekaert$^a$ and S.I. Aadharsh Raj$^{a,b}$}

\date{${}^a$ Institut Denis Poisson, Unit\'e Mixte de Recherche $7013$ du CNRS\\
Universit\'e de Tours, Universit\'e d'Orl\'eans\\
Parc de Grandmont, 37200 Tours, France\\
\vspace{2mm}
{\tt xavier.bekaert@lmpt.univ-tours.fr}\\
\vspace{5mm}
${}^b$ UM-DAE Centre for Excellence in Basic Sciences\\
Mumbai 400098, India\\
\vspace{2mm}
{\tt aadharsh.raj@cbs.ac.in}
}

\maketitle


\begin{abstract}
The relation between two branches of solutions (radiative and subradiative) of wave equations on Minkowski spacetime is investigated, for any integer spin, in flat Bondi coordinates where remarkable simplifications occur and allow for exact boundary-to-bulk formulae.
Each branch carries a unitary irreducible representation of the Poincar\'e group, though an exotic one for the subradiative sector. These two branches of solutions are related by an inversion and, together, span a single representation of the conformal group. While radiative modes are realised in the familiar holographic way (either as boundary data at null infinity or as bulk fields with radiative asymptotic behavior), the whole tower of subradiative modes forms an indecomposable representation of the usual Poincar\'e group, which can be encoded into a single boundary field living on an interior null cone. Lorentz transformations are realised in both cases as conformal transformations of the celestial sphere. The vector space of all subradiative modes carries a unitary representation of a group isomorphic to the Poincar\'e group, where bulk conformal boosts play the role of bulk translations.
\end{abstract}

\thispagestyle{empty}

\pagebreak

\setcounter{page}{0}

\setcounter{tocdepth}{2}

\tableofcontents

\pagebreak

\section{Introduction}

In the sixties, the proper description of gravitational radiation in general relativity \cite{Bondi:1960jsa,Bondi:1962px,Sachs:1962wk} and the concomitant discovery of the asymptotic infinite-dimensional enhancement of symmetries in this context \cite{Bondi:1962px,Sachs:1962wk,Sachs:1962zza} prompted systematic investigations of the conformal completion of spacetimes \cite{Penrose:1962ij,Penrose:1965am,Geroch:1977big}. Later on, the asymptotic quantization program \cite{Ashtekar:1987tt} relied on the systematic investigation of the encoding of radiative modes at null infinity. More recently, the extension of Bondi-Metzner-Sachs symmetries by super-rotations \cite{Barnich:2009se,Barnich:2010ojg} and the discovery of the infrared triangle \cite{Strominger:2017zoo} as well as the ongoing quest of ``flat holography'', i.e. holography for asymptotically flat spacetimes (see e.g. the recent reviews \cite{Pasterski:2021raf} on celestial holography), triggered a surge of interest on the asymptotic structures (charges, modes, etc) of gravity and gauge theories.
These motivate the present systematic investigation of the radiative and subradiative solutions of free wave equations for fields of any integer spin on Minkowski spacetime.

Being second order, it is natural that wave equations admit two branches of solutions. On the one hand, this is a standard feature of holographic investigations around asymptotically anti de Sitter (AdS) spacetimes.
On the other hand, most holographic investigations around asymptotically flat spacetimes focus on radiative modes for natural physical reasons (since electromagnetic and gravitational radiation is of utmost importance). Nevertheless, the sector of subradiative modes is expected to be relevant in flat holography, as emphasised for instance in \cite{Mittal:2022ywl} where subradiative modes were dubbed ``chthonian'' (since they arise at orders \textit{below} the radiative order). One of our goal is to show that, for conformally-covariant wave equations (such as d'Alembert, Maxwell and Bargmann-Wigner equations), the radiative and subradiative sectors are intimately related by an inversion. This property is another hint that both sectors are of importance and that subradiative modes should not be disregarded.

Throughout the paper, ``flat'' Bondi coordinates will be used extensively because illuminating simplifications occur in these coordinates. Recall that this coordinate system is constructed in a similar way to usual Bondi coordinates (which will be refered in the present work as ``round'' Bondi coordinates, to avoid confusion) except that the flat metric is used as representative of the conformal metric on the celestial sphere (deprived of a point). The fact that technical simplifications happen in these alternative Bondi coordinates is known for some time \cite{Barnich:2009se,Barnich:2010eb} for asymptotically flat spacetimes. The asymptotic coordinate transformation allowing to go from the round to the flat representative of the conformal metric on the celestial sphere was provided in \cite{Barnich:2016lyg}. 
More recently, the exact coordinate transformation on Minkowski spacetime between Cartesian and flat Bondi coordinates have been constructed and used in the Carrollian approach to flat holography in \cite{Donnay:2022wvx,West}.
A remarkable example of the simplifications occuring in flat Bondi coordinates that we point out is that the stationary phase approximation becomes exact (in the sense that it is a Gaussian integral which can of course be performed explicitly) and provides an exact formula between boundary data and bulk solution, in analogy with the AdS/CFT dictionary.
Another remarkable simplification in flat Bondi coordinates is that the retarded time and (the inverse of) the radial coordinate on Minkowski spacetime are exchanged under an inversion. Consequently, in these coordinates the duality between the radiative and subradiative sector is totally manifest. 
Geometrically, the inversion exchanges an interior null cone with the null cone at infinity. Accordingly, the duality between the two asymptotic expansion schemes (near null infinity and near an interior null cone) is also manifest.

A summary of our results and plan of the paper is presented in the next subsection, which is followed by another subsection where conventions and notations are presented.

\subsection{Plan of the paper}

In Section \ref{dalembert}, we treat in many details the example of the massless scalar field, since it is technically simpler and serves as an illuminating illustration of the general case.
In particular, we expand the solutions of d'Alembert equation around an interior null cone and around null infinity, and explain how these two expansion schemes are related to each other via an inversion. We also generalise the stationary phase approximation relating the asymptotic behaviour of the bulk solution to the boundary data at null infinity to an exact reconstruction formula from the boundary to the bulk.
In Section \ref{Max}, we show that Maxwell equations (in four space-time dimensions) in flat Bondi coordinates are manifestly invariant under the inversion duality exchanging radiative and subradiative solutions.
In Section \ref{anyspin}, we show that these features generalise to any integer spin for Bargmann-Wigner equations. Finally, in Section \ref{concl} we briefly summarise our findings and mention few directions for the future. Three appendices follow.

\subsection{Conventions}

We will work in any spacetime dimension $D\geqslant 2$. More precisely, we will consider Minkowski spacetime of dimension $D=d+2$, denoted $\mathbb{R}^{d+1,1}$. The celestial sphere is of dimension $d\geqslant 0$. Future/past null infinity are denoted $\mathscr{I}^\pm_{d+1}\cong\mathbb{R}\times S^d$.
Greek indices $\mu,\nu,\rho,\ldots$ will run from $0$ to $d+1$ while Latin indices $i,j,k,\ldots$ will run from $1$ to $d$.
    The boldface symbols $\mathbf x$ and $\mathbf y$ will denote points of the Euclidean plane $\mathbb{R}^d$ with respective Cartesian coordinates $x^i$ and $y^i$.
The signature of the metric is mostly plus: $\eta_{\mu\nu}=\text{diag}(-1,+1,\ldots,+1)$.
The connected component of the Lorentz group $SO(d+1,1)$ will be denoted
$SO_0(d+1,1)$.
The connected group $ISO_0(d+1,1)=SO_0(d+1,1)\ltimes\mathbb{R}^{d+1,1}$ will be called the Poincar\'e group.
The conformal group is $O(d+2,2)/\mathbb{Z}_2$\,, where $\mathbb{Z}_2=\{+I,-I\}$.
The connected component of the full conformal group $O(d+2,2)/\mathbb{Z}_2$ will be denoted $SO_0(d+2,2)$ and called the restricted conformal group. 

\section{Massless scalar field}\label{dalembert}

\subsection{Bondi coordinates on Minkowski spacetime: flat vs round}\label{subsec: Bondi Coordinates}

We provide a quick review, based on the appendix A in \cite{Donnay:2022wvx} (generalised to higher dimensions in \cite{West}), of the flat and round Bondi coordinates on Minkowski spacetime, since they are instrumental in our investigations. In Subsection \ref{flatBondicords}, we will point out a Weyl transformation that dramatically simplifies computations for wave equations invariant under conformal transformations and makes manifest the inversion duality between the radiative and subradiative sectors.

\subsubsection{Round Bondi Coordinates}

For Minkowski spacetime, the usual Bondi coordinates coincide with the Eddington-Finkelstein coordinates. The Cartesian coordinates $x^\mu$ on $\mathbb{R}^{d+1,1}$ can be expressed in terms of the retarded Bondi coordinates $(\tilde{r},\tilde{u},\mathbf{x})$ as follows:
\begin{equation}\label{BondiCart}
x^{\mu} = \Tilde{u}\:\Tilde{n}^{\mu} + \Tilde{r}\:\Tilde{q}^{\mu} \,,   
\end{equation}
where the vector fields $\Tilde{n}$ and $\Tilde{q}$ obey the following 3 conditions:
\begin{enumerate}
    \item $\Tilde{n}$ is a constant unit timelike vector ($\Tilde{n}^2=-1$) $\implies \Tilde{n}^{\mu} = \delta^{\mu}_0$ (without loss of generality).
    \item $\Tilde{q} \cdot \Tilde{n} = -1 \implies \Tilde{q}^0 = 1$.
    \item $\Tilde{q}$ is null ($\Tilde{q}^2 = 0$) $\implies \Tilde{q}^{\,i}\Tilde{q}_{\,i} = 1$.
\end{enumerate}
The last condition implies that the vector field $\Tilde{q}^{\,i}(\theta^j)$ can be parameterised using spherical coordinates $\theta^j$ on the unit sphere. These three condition imply that $x^i=\Tilde{r}\:\Tilde{q}^{\,i}$, where $\Tilde{r} = \sqrt{x^ix_i}$ is the usual radial coordinate. Therefore, the coordinate $\Tilde{u} = {t} - \Tilde{r}$ is the retarded time. The line element of Minkowski spacetime reads 
\begin{equation}
ds^2_{{}_{\mathbb{R}^{d+1,1}}} = -d{t}^2 + d\Tilde{r}^2 + \Tilde{r}^2 d\Omega^2 = -d\Tilde{u}^2 - 2\,d\Tilde{u}\, d\Tilde{r} + \Tilde{r}^2 d\Omega^2     
\end{equation}
where $d\Omega^2$ is the line element on the unit sphere $S^d$. There is a  coordinate singularity at $\tilde r=0$, whose locus is the time-like straight line $x^{\mu} = \Tilde{u}\:\Tilde{n}^\mu$, cf. \eqref{BondiCart}. 

\subsubsection{Flat Bondi Coordinates}\label{flatBondicords}

We now review the slight modification of the 1st condition that allows to obtain ``flat'' Bondi coordinates  \cite{Donnay:2022wvx,West}. 

Before that we introduce light-cone coordinates, as we will be playing around with many null vectors below.
\begin{equation}\label{lightconmetric}
x^{\pm} = \frac{1}{\sqrt{2}}\left(x^0 \pm x^{d+1} \right) \implies ds^2_{{}_{\mathbb{R}^{d+1,1}}} = -2\,dx^+dx^- + dx^idx_i\,.    
\end{equation}
The expression of the Cartesian coordinates in terms of the (retarded) flat Bondi coordinates $(u,r,y^i)$ take the same form as \eqref{BondiCart}:
\begin{equation}
\label{coord_trans}
    x^{\mu} = u\:n^{\mu} + r\:q^{\mu}
\end{equation}
Now the modified conditions are:
\begin{enumerate}
    \item $n$ is constant and null ($n^2=0$) $\implies n^{\mu} = \delta^{\mu}_{-}$ (can be taken normalised and along $x^-$ direction without loss of generality)
    \item $q \cdot n = -1 \implies q^+ = 1$.
    \item $q$ is null ($q^2 = 0$) $\implies 2\,q^- = q^iq_i$.
\end{enumerate}
We normalise as follows the coordinates $y^i :=  q^i/\sqrt{2}$ on the Euclidean plane $\mathbb{R}^d$ in order to align with the conventions in \cite{West}. Then 
\begin{equation}\label{coord_transq}
q^{\mu}(\mathbf y) = (q^+,q^-,q^i) = (1,\abs{y}^2, \sqrt{2}\,y^i)\,.    
\end{equation}
This defines our flat Bondi coordinates. We can work out the expression of the norm squared of the null vector field $q$ at two distinct points:  
\begin{equation}\label{scalarprodq}
q^\mu(\mathbf x)\,q_\mu(\mathbf y) = -\,q^+(\mathbf x)\,q^-(\mathbf y) - \,q^-(\mathbf x)\,q^+(\mathbf y) + q^i(\mathbf x)\, q_i(\mathbf y)  = - \abs{\mathbf x}^2 - \abs{\mathbf y}^2 + 2\, {\mathbf x}\cdot{\mathbf y} = -\abs{\mathbf x -\mathbf  y}^2\,.    
\end{equation}
In the flat Bondi coordinates $(u,r,y^i)$, the metric takes the very simple form
\begin{equation}
    \label{metric}
    ds^2_{{}_{\mathbb{R}^{d+1,1}}} = -2\, du\, dr + 2\,r^2 dy^idy_i
\end{equation}
with a coordinate singularity at $r=0$ (similarly to the one for round Bondi coordinates, with the important difference that here the locus of this coordinate singularity is a codimension one hypersurface, see below). 
Geometrically, $u$ is the value of the affine time on the null geodesic along the vector $n^\mu$, at which another null geodesic is emitted in the direction $q^\mu$.
With a slight abuse of terminology, the coordinate $u$ will be called the ``retarded time'' and $r$ the ``radial coordinate''. 

We can invert the relation \eqref{coord_trans} to give flat Bondi coordinates in terms of light-cone coordinates,
\begin{equation}
\label{inv_trans}
r = x^+ \,, \quad u = x^- - \frac{x^ix_i}{2x^+} \,,\quad y^i = \frac{x^i}{\sqrt{2}\,x^+}\,. \end{equation}
The locus of the coordinate singularity is the null hyperplane $x^+ = 0$, where the relations \eqref{inv_trans} break down. In fact, the flat Bondi coordinates are not globally defined: they only cover the patches $x^+>0$ ($r>0$) and $x^+<0$ ($r<0$). Future and past null infinity $\mathcal{I}_{d+1}^\pm$ correspond to the limit $r \to \pm\infty$ in the corresponding patch. In particular, the coordinates $(u,y^i)$ only cover a part of null infinity because a null line is excluded (corresponding to the endpoints of the null geodesics parallel to $n$). Equivalently, the point at infinity is excluded from the celestial sphere for each value of retarded time.\footnote{That is to say: 
$\mathcal{I}_{d+1}^\pm\backslash \mathbb{R}\cong \mathbb{R}\times \mathbb{R}^d$ (since $\mathcal{I}_{d+1}^\pm\cong\mathbb{R}\times S^d$ and
$S^d\backslash\mathbb{R}\cong\mathbb{R}^d$). See also p.155 and Fig.6 of \cite{Penrose}.} With this picture in mind, the plane $\mathbb{R}^d$ with coordinates $y^i$ will sometimes be referred to as the ``celestial plane'' (since it corresponds to the celestial sphere, with one point excluded). 

The definition \eqref{coord_trans}, and the fact that the vectors $n$ and $q$ form a null dyad, imply that the norm squared of the position vector takes the simple form $x^2=-2ur$. As one can see, the equation $x^2=0$ for the null cone $\mathcal{N}_{d+1}$ through the origin takes the simple form $u=0$ in the above flat Bondi coordinates.\footnote{\label{ur=0}Note that $r=0$ is the equation of the null line $x^\mu = u n$, which is the intersection between the hyperplane $x^+=0$ and the null cone $x^2=0$.} The null cone through the origin will often be referred to as ``the null cone'' for short, since it will play a prominent role in our analysis.
The intersection, $\mathscr{N}_{d+1}^+\cap\mathscr{I}_{d+1}^+\cong S^d$, between the future null cone through the origin and future null infinity is the good cut at null infinity that corresponds to the origin of Minkowski spacetime. Up to one missing point, in flat Bondi coordinate the system of two equations for this cut are $u=0$ and $r=\infty$.

The usual inversion $x^\mu\to\frac{x^\mu}{x^2}$ takes the simple form $u\to -\tfrac1{2r}$ and $r\to -\tfrac1{2u}$ in flat Bondi coordinates since
\begin{equation}
x^{\mu} \to \frac{x^\mu}{x^2}=\frac{un^{\mu} + rq^{\mu}}{-2ur} =  -\frac{1}{2} \left(\frac{1}{r}n^{\mu} + \frac{1}{u}q^{\mu}\right)\,.
\end{equation}
This composition of this inversion with a parity transformation $x^\mu\to -x^\mu$, followed by the dilatation $x^\mu\to 2 x^\mu$, is the conformal transformation  $x^\mu\to-\frac{2x^\mu}{x^2}$. Due to the choice of normalisation $n\cdot q=-1$, it is the latter transformation that takes the simplest form $u\leftrightarrow \tfrac1{r}$. With a slight abuse of terminology (since it involves a parity transformation), we will refer to this conformal transformation as an inversion.

To conclude, we mention a last curiosity of flat Bondi coordinates, which proves to be extremely useful for our purpose: the retarded time and the inverse radius $v:=1/r$ play the role of (rescaled) light-cone coordinates for a flat metric, $d\tilde s^2 = du\, dv + dy^idy_i$\,, conformally related to the original metric on Minkowski spacetime. Indeed, in terms of the coordinates $(u,v,y^i)$, the metric \eqref{metric} takes the very simple form
\begin{equation}
\label{metric2}
ds^2_{{}_{\mathbb{R}^{d+1,1}}} = \frac2{v^2}\,(du\, dv + dy^idy_i)
\end{equation}
since $dv=-\frac{dr}{r^2}$. Therefore, the Minkoswki metric \eqref{metric2} is related by a Weyl transformation $ds^2\to d\tilde s^2=\Omega^2ds^2$ with conformal factor $\Omega=\frac{|v|}{\sqrt{2}}$ to the light-cone metric $d\tilde s^2 = du\, dv + dy^idy_i$, cf. \eqref{lightconmetric} with $x^+=u$ and $x^-=-v/2$. As will be made manifest later, this simple observation allows to simplify substantially the asymptotic analysis in flat Bondi coordinates for wave equations with Weyl symmetry (such as massless fields of any spin in four-dimensional Minkowski spacetime) because the above Weyl transformation allows to map the original problem (the holographic description of bulk solutions in terms of boundary data at null infinity and on the interior null cone) to a simpler problem in light-cone coordinates (a higher-dimensional version of Goursat problem, where the boundary data is fixed on two intersecting null hyperplanes).

\subsection{Expansion near null infinity}\label{expnullinfty}

We now look at the wave equation for a massless scalar field $\upphi$ on Minkowski spacetime $\mathbb{R}^{d+1,1}$. As we are interested in the asymptotic behaviour near $\mathcal{I}^+_{d+1}$, we assume an expansion in powers of $1/r$:
\begin{equation}\label{asexp1/r}
\upphi(u,r,\mathbf y) = \frac{1}{r^{\Delta}}\sum_{n=0}^{\infty} \frac{1}{r^{n}} \,\phi_n(u,\mathbf y)
\end{equation}
where the scaling dimension $\Delta\in\mathbb R$ can, in full generality, take any real value.
The d'Alembertian reads in flat Bondi coordinates
\begin{equation}\label{dAlembert}
\Box_{{}_{\mathbb{R}^{d+1,1}}} =-\frac{d}{r}\, \partial_u - 2\,\partial_r\partial_u + \frac{1}{2r^2}\, \partial_i\partial^i    
\end{equation}
where $\partial_i\partial^i$ is the Laplacian on the Euclidean plane $\mathbb{R}^d$.
The ansatz \eqref{asexp1/r} implies that the d'Alembert equation $\Box_{{}_{\mathbb{R}^{d+1,1}}}\upphi =0$ amounts to the following infinite system of equations on the coefficients in the asymptotic expansion:
\begin{equation}\label{system1/r}
\left(-d + 2\Delta + 2n \right) \partial_u \phi_n + \frac{1}{2} \partial_i\partial^i  \phi_{n-1} = 0\,,\qquad (n=0,1,2,\ldots)      
\end{equation}
where $\phi_{-1} = 0$. For $n = 0$, we get,
\begin{equation}
\left(-d + 2\Delta \right) \partial_u \phi_0   = 0      \,.
\end{equation}
There are two distinct cases to be discussed: either $\Delta = \frac{d}{2}$ and the leading coefficient $\phi_0$ is left arbitrary, or $\Delta \neq \frac{d}{2}$ and the leading coefficient $\phi_0$ must be independent of the retarded time. 

Let us focus on the first case from now on, i.e. we set $\Delta = \frac{d}{2}$.\footnote{See e.g. \cite{Satishchandran:2019pyc,Bekaert:2022ipg,Bekaert:2024itn} for a systematic discussion of the case $\Delta \neq \frac{d}{2}$ in round Bondi coordinates.}
In other words, we focus on the solution space of the wave equation with asymptotic behaviour \eqref{asexp1/r} for $\Delta = \frac{d}{2}$, in particular
\begin{equation}\label{subspaceasympt1}
\Box_{{}_{\mathbb{R}^{d+1,1}}}\upphi =0\,,\qquad \upphi(u,r,\mathbf y)\stackrel{r\to\infty}{\sim}\frac{1}{r^{\frac{d}2}}\,\Big[\,\phi_0(u,\mathbf y)+\mathcal{O}\left(\tfrac1{r}\right)\,\Big]\,.
\end{equation}
The solutions $\upphi$ with $\phi_0\neq 0$ will be called \textit{radiative modes}, while the solutions with 
$\phi_0= 0$ will be called \textit{subradiative modes}. More generally, solutions of the problem \eqref{subspaceasympt1} will be called \textit{radiative solutions} (whether $\phi_0$ vanishes or not). It is known that solutions of the Cauchy problem with smooth initial data
of compact support are radiative solutions (see e.g. the discussion in \cite[Section II.A]{Satishchandran:2019pyc}). 

The vector space of solutions to the d'Alembert equation with asymptotic expansion \eqref{asexp1/r} is an infinite-dimensional $\mathfrak{iso}(d+1,1)-$module, which will be denoted
\begin{equation}
\label{Viso}
\mathcal{V}^{\mathfrak{iso}(d+1,1)}(\Delta,0)\,:=\,\bigg\{\upphi(u,r,\mathbf y) = \frac1{r^{\Delta}} \sum_{n \geqslant 0} \frac1{r^{n}}\,\phi_n(u,\mathbf y)\,:\,\Box_{{}_{\mathbb{R}^{d+1,1}}}\upphi=0\bigg\}\,. 
\end{equation}
In particular, $\mathcal{V}^{\mathfrak{iso}(d+1,1)}(\tfrac{d}{2},0)$ is the vector space of radiative solutions while $\mathcal{V}^{\mathfrak{iso}(d+1,1)}(\tfrac{d}{2}+1,0)$ is the vector space of subradiative modes.
The latter is a Poincar\'e-invariant subspace of the former, $\mathcal{V}^{\mathfrak{iso}(d+1,1)}(\tfrac{d}{2}+1,0)\subset\mathcal{V}^{\mathfrak{iso}(d+1,1)}(\tfrac{d}{2},0)$.
In fact, the action of Poincar\'e generators (in particular translation generators) is triangular on the collection of coefficients $\phi_n$, for instance $\hat{P}_\mu\phi_n$ contains a contribution from $\phi_{n-1}$ since $\partial_{\mu}=q_\mu(\mathbf y)\partial_u+\mathcal{O}(\tfrac1{r})$.

For $\Delta = \frac{d}{2}$ and $n \geqslant 1$, the equations \eqref{system1/r} take the simple form
\begin{equation}\label{systemd/2}
\partial_u \phi_n + \frac{1}{4n}\,\partial_i\partial^i  \phi_{n-1}=0\,.
\end{equation}
Since the action of Poincar\'e generators, as well as the system \eqref{systemd/2} of equations, have a triangular structure, the $\mathfrak{iso}(d+1,1)-$module $\mathcal{V}^{\mathfrak{iso}(d+1,1)}(\tfrac{d}{2},0)$ of radiative solutions is reducible but not fully reducible, i.e. it is indecomposable. In other words, there is no Poincar\'e-invariant decomposition of radiative solution between ``pure'' radiative modes and subradiative modes. Nevertheless, as we will see below, if one makes a choice of origin in Minkowski spacetime, then there exists such a decomposition (but this is decomposition will clearly not be translation-invariant, by construction).
We will sometimes informally use the terms radiative and subradiative sectors, though it is slightly improper (since, strictly speaking, the subradiative modes form a \textit{subsector} of the radiative sector).

We can solve the differential equation \eqref{systemd/2} of first-order in retarded time by a direct integration, that is to say
\begin{equation}\label{phin}
\phi
_n (u,\mathbf y)= - \frac{1}{4n} \int_0^{u} du' \,\partial_i\partial^i  \phi_{n-1}(u',\mathbf y)\,+\, \psi_n(\mathbf y)    \,,
\end{equation}
where $\psi_n(\mathbf y):=\phi_n (u=0,\mathbf y)$ for $n\geqslant 1$ are ``integration constants'', which are arbitrary functions on the celestial plane. (Note that we implicitly assumed that all coefficients $\phi_n (u=0,\mathbf y)$ have a smooth behaviour in the limit $u\to 0$.) 
The infinite tower of these ``integration constants'' $\psi_n(\mathbf y)$ can be packaged into a single generating function
\begin{equation}\label{Psi0}
\psi_0(r,\mathbf y)\,:=\,
\sum_{n = 1}^{\infty} \frac{1}{r^{n}} \,
\psi_n(\mathbf y)\,=\,r^{\frac{d}2}\upphi(u = 0, r,\mathbf y)\,-\,\phi_{0}(u = 0, \mathbf y)\,.
\end{equation}
This generating function is a function on the null cone $\mathcal{N}_{d+1}$ through the origin. 
One can try to summarise this observation in group-theoretical terms by saying that the submodule of subradiative modes is isomorphic to the vector space of boundary data on the null cone, i.e. 
\begin{equation}\label{isomorphNullcone}
\mathcal{V}^{\mathfrak{iso}(d+1,1)}(\tfrac{d}{2}+1,0)\cong C^\infty(\mathcal{N}_{d+1})\,.    
\end{equation}

Substituting now the expression for $\phi_{n-1}$ in the previous equation \eqref{phin} gives
\begin{equation}
\phi_n =  \frac{1}{4^2n(n-1)}\int_0^{u} du'\int_0^{u'} du''  (\partial_i\partial^i)^2 \phi_{n-2}(u'',\mathbf y) - \frac{u}{4n}\,  \partial_i\partial^i
\psi_{n-1}(\mathbf y) + \psi_n(\mathbf y)    
\end{equation}
and so on. 
The general solution can be written explicitly in a compact form, as will be done in Section \ref{reconstructionform} and Appendix \ref{heuristic}, but its structure is already clear: the general solution $\upphi$ is a sum of a leading piece $\upphi_{\text{rad}}$, which is a radiative solution entirely determined by the boundary data $\phi_0(u,\mathbf y)$ at null infinity, plus a subradiative solution $\upphi_{\text{sub}}(u,r,\mathbf y)$ which is entirely determined by the boundary data $\psi_0(r,\mathbf y)$ on the null cone.

What we called above the ``leading piece'' is sometimes refered to as ``radiative'' or ``radiation'' field (for $s>0$) in the literature (see e.g. \cite{Adamo:2021dfg}). With a slight abuse of terminology, a radiative solution with asymptotic expansion \eqref{asexp1/r} where $\Delta=\tfrac{d}{2}$, which is such that 
$\phi_{n}(u=0,\mathbf y)=0$, for all positive integers $n\geqslant 1$, will be called a \textit{radiation field}:
\begin{equation}
\upphi_{\text{rad}}(u,r,\mathbf y) = \frac{1}{r^{\tfrac{d}{2}}}\sum_{n=0}^{\infty} \frac{1}{r^{n}} \,\phi_n(u,\mathbf y)\;\;\text{such that}\;\;\Box_{{}_{\mathbb{R}^{d+1,1}}}\upphi_{\text{rad}}=0\;\;\text{and}\;\;\phi_{n}(u=0,\mathbf y)=0, \;\;\forall n\geqslant 1\,. 
\end{equation}
Equivalently, a radiation field is a radiative solution such that $\psi_0(r,\mathbf y)=0$. Due to \eqref{Psi0}, this condition is also equivalent to the condition
$r^{\tfrac{d}{2}}\upphi_{\text{rad}}(u=0,r,\mathbf y)=\phi_0(u=0,\mathbf y)$. 
The right-hand side, $\phi_0(u=0,\mathbf y)$, is the leading coefficient in the asymptotic expansion, evaluated at the corresponding cut at null infinity. Therefore, a radiation field is a radiative solution whose restriction on the null cone almost vanishes in the sense that its restriction on the null cone $\mathscr{N}_{d+1}^+$ coincides (up to a rescaling) with its further restriction onto the corresponding cut $\mathscr{N}_{d+1}^+\cap\mathscr{I}_{d+1}^+\cong S^d$ at null infinity. 

The general solution $\upphi$ is a sum 
\begin{equation}
\upphi(u,r,\mathbf y) = \upphi_{\text{rad}}(u,r,\mathbf y)+\upphi_{\text{sub}}(u,r,\mathbf y) \,,
\end{equation}
of a leading piece $\upphi_{\text{rad}}$ and a subleading piece $\upphi_{\text{sub}}$.
The leading piece is a radiation field
\begin{equation}\label{leadingpiece}
\upphi_{\text{rad}}(u,r,\mathbf y)=\frac1{r^{\frac{d}2}}\,\Big[\phi_0(u,\mathbf y)+\,\phi_{\text{tail}}\left(u,r,\mathbf y\right)\Big]\,,
\end{equation}
including a ``tail'', denoted $\phi_{\text{tail}}$\,, which is an $\mathcal{O}(1/r)$ and an $\mathcal{O}(u)$ that contain all the integrals and the Laplacians acting on the boundary data $\phi_0$\,:
\begin{eqnarray}\label{tail}
\phi_{\text{tail}}(u,r,\mathbf y)&=&\sum_{n=1}^{\infty}\frac1{n!}\left(-\frac1{4r}\right)^n\,\int_0^{u} du_1\int_0^{u_{1}} du_2\cdots\int_0^{u_{n-1}} du_n\, (\partial_i\partial^i)^n\phi_0(u_n,\mathbf y)\\
&=&-\,\frac1{4r} \int_0^{u} du' \,\partial_i\partial^i  \phi_0(u',\mathbf y)\,+\,\mathcal{O}\left(\tfrac1{r^2}\right)\,.
\end{eqnarray}
The subleading (hence subradiative) piece takes the form
\begin{equation}\label{subexp}
\upphi_{\text{sub}}(u,r,\mathbf y)=\frac1{r^{\frac{d}2}}\,\Big[\psi_0(r,\mathbf y)+\psi_{\text{tail}}(u,r,\mathbf y)\Big]
\end{equation}
where $\psi_0(r,\mathbf y)=\mathcal{O}(1/r)$ by definition \eqref{Psi0}.
The right-hand side in \eqref{subexp} contains an $\mathcal{O}(1/r^2)$ and $\mathcal{O}(u)$ tail, denoted $\psi_{\text{tail}}$\,, which takes the form 
\begin{eqnarray}\label{tail'}
\psi_{\text{tail}}(u,r,\mathbf y)&=&\sum_{m=1}^{\infty}\frac{\,u^m}{m!}\sum_{n=1}^{\infty}\frac{1}{r^{m+n}}\,\frac{n!}{(m+n)!} \left(-\frac{\partial_i\partial^i}{4}\right)^m\psi_n(\mathbf y)
\\
&=&- \frac{u}{8r^2}\,  \partial_i\partial^i\psi_1(\mathbf y)+\mathcal{O}\left(\tfrac{1}{r^3}\right) \,,
\end{eqnarray}
Note that $r^{\tfrac{d}{2}}\upphi_{\text{sub}}(u=0,r,\mathbf y)=\psi_0(r,\mathbf y)$.

We can now answer the following question: What is the data that must be given in order to determine uniquely the solution $\upphi$ of the wave equation? The answer is, we need to specify two functions (each defined on a distinct null cone, respectively: future null infinity and an interior null cone):
\begin{itemize}
    \item the \textit{radiative data} which is an arbitrary function $\phi_{0}(u,\mathbf{y})$ at future null infinity $\mathcal{I}_{d+1}^+$ and
    \item the \textit{subradiative data} $\psi_{0}(r,\mathbf{y})$ at the future null cone $\mathcal{N}_{d+1}^+$. 
\end{itemize} 
These two data (which are, respectively, the boundary data of the radiative and subradiative modes) can be written as,
\begin{equation}
\label{null_inf_specifics}
\phi_0(u, \mathbf y)=\lim_{r \to \infty} \Big[r^{\frac{d}{2}} \upphi(u, r,\mathbf y)\Big]\,,\qquad \psi_0(r, \mathbf y)=\lim_{u \to 0} \left[r^{\frac{d}2}\upphi(u, r,\mathbf y)\,-\,\phi_{0}(u, \mathbf y)\right]
\end{equation}
Recall that the $\mathfrak{iso}(d+1,1)-$module $\mathcal{V}^{\mathfrak{iso}(d+1,1)}(\tfrac{d}{2},0)$ of radiative modes is indecomposable. Furthermore, the submodule $\mathcal{V}^{\mathfrak{iso}(d+1,1)}(\tfrac{d}{2}+1,0)$ of subradiative solutions is isomorphic to the space of subradiative data $\psi_0$ on the null cone, cf. \eqref{isomorphNullcone}. One can consider  
the quotient module 
\begin{equation}\label{Dirred}
\mathcal{D}^{\mathfrak{iso}(d+1,1)}(\tfrac{d}{2},0)\,:=\,\mathcal{V}^{\mathfrak{iso}(d+1,1)}(\tfrac{d}{2},0)\,/\,\mathcal{V}^{\mathfrak{iso}(d+1,1)}(\tfrac{d}{2}+1,0)\,\cong\, C^\infty(\mathcal{I}_{d+1})
\end{equation} 
of equivalence classes $[\upphi]$ of radiative solutions $\upphi$ modulo subradiative ones $\upphi_{\text{sub}}$. The radiation fields $\upphi_{\text{rad}}$ correspond to a specific choice of representatives for these equivalence classes $[\upphi]$. As written in \eqref{Dirred}, the $\mathfrak{iso}(d+1,1)-$module  $\mathcal{D}^{\mathfrak{iso}(d+1,1)}(\tfrac{d}{2},0)$ is isomorphic to the vector space of radiative data $\phi_0$ at null infinity.\footnote{See also \cite[Sections II.A-B]{Satishchandran:2019pyc} or \cite[Section 3.1]{Bekaert:2024itn} for a systematic discussion of the generic case.} 

Remember that the $\mathfrak{iso}(d+1,1)-$module $\mathcal{V}^{\mathfrak{iso}(d+1,1)}(\tfrac{d}{2},0)$ of radiative solutions is indecomposable. Therefore, this representation of the Poincar\'e group is \textit{not} unitarisable.\footnote{Recall that all unitary representations are fully reducible (simply by taking the orthogonal complement of the invariant subspace). Thence, by contraposition, an indecomposable representation cannot be unitary.}
However, the $\mathfrak{iso}(d+1,1)-$module $\mathcal{D}^{\mathfrak{iso}(d+1,1)}(\tfrac{d}{2},0)$ is unitarisable (if one restricts to a suitable subclass of solutions, e.g. to radiative data $\phi_0$ which are square-integrable at null infinity) and corresponds to the unitary irreducible representation of Wigner corresponding to a massless scalar field. In particular, the usual Wigner hermitian product, of the momentum space (square-integrable) wavefunctions on the null cone, i.e.
\begin{equation}
\langle\upphi \mid \upphi'\rangle:= \int d^{d+2}p \,\delta(p^2) \,\upphi^*(p) \upphi'(p) = \int \frac{d^{d+1} p}{|\overrightarrow{p}|}\, \upphi^*(\overrightarrow{p}) \upphi'(\overrightarrow{p}) 
\quad\text{where}\quad p^\mu=(p^0,\overrightarrow{p}),
\end{equation}
coincides with the Sachs hermitian product \cite{Sachs:1962zza} of (square-integrable) radiative data at null infinity given by
\begin{equation}
\label{sachs_form}
\langle\,\phi_0 \mid \phi'_0\,\rangle =i \int_{-\infty}^{+\infty} d u \int_{\mathbb{R}^d} d^d \mathbf{y} \,\phi_0^*(u,\mathbf{y}) \,\partial_u \phi'_0(u,\mathbf{y})\,.
\end{equation}
This property $\langle\upphi \mid \upphi'\rangle=\langle\,\phi_0 \mid \phi'_0\,\rangle$ will not be rederived here explicitly (see e.g. \cite[Section 2.4]{Bekaert:2022ipg} for details).
Note that space-time translations preserve the above hermitian products as they act on the radiative data at null infinity, as specific shifts in $u$:
\begin{eqnarray}
&&\delta_a\upphi(u,r,\mathbf y)=a^{\mu}\partial_{\mu}\upphi(u,r,\mathbf y) = -\frac1{r^{ \frac{d}{2}}}\, a_\mu q^\mu(\mathbf y)\, \partial_u\phi_0(u,\mathbf y) + \mathcal{O}\left(\tfrac{1}{r}\right)\nonumber\\
&&\implies \delta_a \phi_0(u,\mathbf y)=- \,a_\mu q^\mu(\mathbf y)\, \partial_u\phi_0(u,\mathbf y)
\end{eqnarray}
since $\partial_\mu=-q_\mu\partial_u-n_\mu\partial_r+\mathcal{O}\left(\tfrac{1}{r}\right)$.
Similarly, Lorentz boosts act on the radiative data at null infinity, as conformal transformations of the celestial sphere together with a suitable rescaling of the retarded time by the conformal factor. In fact, the radiative data $\phi_0$ is a Carrollian primary field with precisely the right scaling dimension $\tfrac{d}{2}$ such that \eqref{sachs_form} is invariant (see e.g. \cite{Bekaert:2022ipg} for details). 

\subsection{Expansion near a light-cone}\label{expnullcone}

We look at the massless scalar wave equation, but now for a change, we expand near an interior null cone, say $\mathcal{N}_{d+1}$ the null cone through the origin. More concretely, we assume an expansion of the (rescaled) bulk scalar field in powers of $u$,
\begin{equation}\label{expu}
r^{\frac{d}{2}}\upphi(r,\mathbf y) = u^{\bar\Delta}\sum_{n=0}^{\infty} u^{n} \varphi_n(r,\mathbf y)\,,
\end{equation}
where $\bar\Delta\in\mathbb R$.
For $\bar\Delta=0$, the solutions of d'Alembert equation which are of order $u$ (i.e. $\varphi_0=0$) will be said \textit{evanescent modes}, since they vanish on the null cone $u=0$.

The expression of the wave operator \eqref{dAlembert} and the ansatz \eqref{expu} imply that d'Alembert equation $\Box_{{}_{\mathbb{R}^{d+1,1}}}\upphi =0$ reduces to the following infinite system of equations on the coefficients $\varphi_n$ in the expansion,
\begin{equation}\label{infsystu}
- 2(\bar\Delta + n) \partial_r\varphi_n + \frac{1}{2r^2} \partial_i\partial^i  \varphi_{n-1}=0
\end{equation}
where $\varphi_{-1} = 0$. For $n = 0$, we get,
\begin{equation}
\bar\Delta\, \partial_r \varphi_0 = 0    
\end{equation}
Again there are two cases: the case $\bar\Delta = 0$, which leaves $\varphi_0(r,\mathbf y)$ arbitrary, or the case $\Delta \neq 0$, which imposes that $\varphi_0$ is independent of $r$. We will focus on the first case $\bar\Delta = 0$ (because we will see that it is the one compatible with the case $\Delta = d/2$ in the $1/r$ expansion). 
In other words, we will focus on the space of radiative solutions to the wave equation that have a finite restriction to the null cone $u=0$, 
\begin{equation}\label{subspaceasympt2}
\Box_{{}_{\mathbb{R}^{d+1,1}}}\upphi =0\,,\qquad \upphi(u,r,\mathbf y)\stackrel{u\to 0}{\sim}\,\frac1{r^{\frac{d}{2}}}\Big[\,\varphi_0(r,\mathbf y)+\mathcal{O}(u)\,\Big]\,.
\end{equation}
More precisely, we will focus on the space of solutions to d'Alembert equation with radiative behaviour near null infinity and with smooth behaviour near the null cone, in the sense of \eqref{subspaceasympt1} and \eqref{subspaceasympt2}. The compatibility between these two behaviours on the cut $(r,u,\mathbf y)=(\infty,0,\mathbf y)$ at infinity implies the ``continuity condition'' 
\begin{equation}\label{subspaceasympt3}
\lim_{u\to 0} \phi_0(u,\mathbf y)=\lim_{r \to \infty} \varphi_0(r,\mathbf y)=f(\mathbf y)\,,
\end{equation}
where $f(\mathbf y)$ is a function on the celestial plane $\mathbb{R}^d$.

For $\bar\Delta=0$ and $n \geqslant 1$, the system \eqref{infsystu} becomes 
\begin{equation}\label{systeqs}
\partial_r\varphi_n  - \frac{1}{4nr^2}\, \partial_i\partial^i  \varphi_{n-1} = 0    
\end{equation}
Integrating this equation, one finds
\begin{equation}
\varphi_n(r,\mathbf y) = \frac1{4n}\int_{\infty}^r \frac{dr'}{r'^2} \, \partial_i\partial^i  \varphi_{n-1}(r',\mathbf y) \,+\, \chi_n(\mathbf y)\,,
\end{equation}
where $\chi_n$ are ``integration constants'' similar to the $\psi_n$ in the asymptotic expansion near null infinity.
Substituting in $\varphi_{n-1}$ leads to
\begin{equation}
\varphi_n(r,\mathbf y) = \frac{1}{4^2n(n-1)}\int_{\infty}^r  \frac{dr'}{r'^2}  \int_{\infty}^{r'} \frac{dr''}{r''^2}  (\partial_i\partial^i)^2\varphi_{n-2}(r'',\mathbf y)\,-\, \frac{1}{4nr}\,\partial_i\partial^i \chi_{n-1}(\mathbf y)  \,+\,  \chi_n(\mathbf y)
\end{equation}
and so on. 
We can see that the structure of recursion relation are essentially the same, with the replacements $u \leftrightarrow \frac{1}{r}$ together with $\phi_n\leftrightarrow \varphi_n$ and $\psi_n\leftrightarrow \chi_n$. Accordingly, the structure of the general solution is again clear: the general solution \eqref{expu} of the wave equation with boundary conditions \eqref{subspaceasympt3} is a sum of two terms. The first term is the leading piece near the null cone, which is entirely determined by the boundary data $\varphi_0(r,\mathbf y)$ on the null cone. The second term is an evanescent solution, entirely determined by the ``integrating constants'' $\chi_n(\mathbf y)$, which can be packaged into the generating function 
\begin{equation}
\label{chi0}
\chi_0(u, \mathbf y):=\sum_{n=1}^{\infty} u^{n} \chi_n(\mathbf y)\,,    
\end{equation}
which can be interpreted as a function at null infinity.

To summarise, the boundary data that we need to specify in order to have a unique solution consists again in two arbitrary functions, one on the interior null cone and one at null infinity,
\begin{equation}
\label{null_cone_specifics}
\varphi_0(r, \mathbf y)=\lim_{u \to 0} \Big[r^{\frac{d}2}\upphi(u, r,\mathbf y)\Big]\,,\qquad \chi_0(u, \mathbf y)=\lim_{r \to \infty} \left[r^{\frac{d}2}\upphi(u, r,\mathbf y)\,-\,\varphi_{0}(r, \mathbf y)\right]\,.
\end{equation}
We refer the reader to Appendix \ref{regularity} for some comments on the fact that the boundary data contains two independent functions. We also briefly discuss the (ir)regularity of solutions in the deep interior of Minkowski spacetime.

Translations do not preserve the null cone through the origin. Relatedly, translations somewhat mess up the expansion \eqref{expu} in power of $u$ (in fact, translations shift the coordinate $u$).
However, special conformal transformations of (compactified) Minkowski spacetime preserve the null cone and, thus are more natural with respect to the expansion \eqref{expu}. In fact, 
special conformal transformations act on the leading modes $\varphi_0$ on the null cone via specific shifts in the inverse radius $v=\tfrac1{r}$. More explicitly\footnote{To show this relation in a covariant way, one can use the following relations:
\begin{equation}
b_\mu x^\mu=b_\mu n^\mu\,u+b_\mu q^\mu\,r=(b_\mu q^\mu)\,r+\mathcal{O}(u)\,,\quad x^\mu\partial_\mu=u\partial_u+r\partial_r=r\partial_r+\mathcal{O}(u)\,,\quad x^2=-2ur=\mathcal{O}(u)\,,
\end{equation} 
that follow from the definition \eqref{coord_trans} and the 3 conditions on the vectors $n$ and $q$.}
\begin{eqnarray}
&&\delta_b\upphi(u,r,\mathbf y)=\,b^{\mu}\left(x_\mu x^\nu \partial_\nu-\frac12\,x^2 \partial_\mu\right)\upphi(u,r,\mathbf y) = -\frac1{r^{ \frac{d}{2}}}\, b_\mu q^\mu(\mathbf y)  \,\partial_v \varphi_0(v,\mathbf y) + \mathcal{O}(u)  \nonumber\\
&&\implies \delta_b \varphi_0(v,\mathbf y)= - b_\mu q^\mu(\mathbf y)  \,\partial_v \varphi_0(v,\mathbf y)\,.
\end{eqnarray}
These special conformal transformations can be checked to preserve the following hermitian product between leading modes on the light-cone
\begin{equation}
\label{sachs_form2}
\langle\,\varphi_0 \mid \varphi_0'\,\rangle =i \int_{-\infty}^{+\infty} d v \int_{\mathbb{R}^d} d^d \mathbf{y} \,\varphi_0^*(v,\mathbf{y})\, \partial_v \varphi'_0(v,\mathbf{y})\,.
\end{equation}
This is obvious since, to switch from \eqref{sachs_form} to \eqref{sachs_form2}, we merely replaced $u$ by $v$.
With respect to Lorentz boosts, the leading modes $\varphi_0$ near the null cone transform as Carrollian primary field of scaling dimension $\tfrac{d}{2}$, which ensures the invariance of the hermitian form \eqref{sachs_form2} under Lorentz boosts.
Let $ISO_0(d+1,1)$ denote the Poincar\'e group generated by translations and boosts.
The Lorentz boosts and the special conformal transformations generate a group $\widetilde{ISO}_0(d+1,1)$ isomorphic to the usual Poincar\'e group  $ISO_0(d+1,1)$.
In this sense, the vector space of leading modes $\varphi_0$ near the null cone provides an ``exotic'' unitary irreducible representation of Poincar\'e group. More accurately, the leading modes carry a unitary irreducible representation of an ``exotic'' Poincar\'e subgroup $\widetilde{ISO}_0(d+1,1)\subset SO_0(d+2,2)$ of the restricted conformal group. 

To complete the analogy with the asymptotic analysis in the previous section, one notices that the vector space of boundary data $\varphi_0$ on the null cone is isomorphic to the vector space of radiative solutions of d'Alembert equation modulo evanescent ones, hence the latter space carries a unitary irreducible representation of the exotic Poincar\'e group $\widetilde{ISO}_0(d+1,1)$. We will comment more on this at the end of the next subsection (see also Section \ref{groupth}).

\subsection{Inversion duality}\label{invduality}

We can now relate the expansions near the null-cone and near null infinity. Using the data \eqref{null_inf_specifics} and \eqref{null_cone_specifics} together with the continuity condition \eqref{subspaceasympt3}, we can conclude that the explicit relations are 
\begin{equation}
\lim_{u \to 0} \Big[r^{\frac{d}{2}}\upphi(u, r,\mathbf y)\Big]=\varphi_0(r, \mathbf y)=\psi_0(r, \mathbf y)+f(\mathbf y)
\end{equation}
and
\begin{equation}
\lim_{r \to \infty} \Big[r^{\frac{d}{2}} \upphi(u, r,\mathbf y)\Big]=\phi_0(u, \mathbf y)=\chi_0(u, \mathbf y)+f(\mathbf y)\,.
\end{equation}
The compatibility with \eqref{subspaceasympt3} is guaranteed by the properties that 
\begin{equation}
\lim_{r \to \infty}\psi_0(r, \mathbf y)=0\,,\qquad\lim_{u \to 0}\chi_0(u, \mathbf y)=0\,,
\end{equation}
which are in agreement with the definitions \eqref{Psi0} and \eqref{chi0}.
We can state the one-to-one correspondence between modes at null infinity and modes on the null cone as follows: \textit{subleading modes $\chi_0$ on the null cone correspond to leading modes $\phi_0$ at null infinity} and \textit{subleading modes $\psi_0$ at null infinity correspond to leading modes $\varphi_0$ on the null cone}. 

One explanation for this phenomenon is that the leading and subleading modes are actually related by an inversion with respect to the origin of the null cone. This is somewhat natural since, geometrically, this inversion (roughly\footnote{See e.g. \cite{Bekaert:2022ipg} (Figure 1 and Footnote 8) for more detailed statements.}) exchanges the two relevant null cones: the one through the origin (i.e. the null cone $\mathscr{N}_{d+1}$) with the one through timelike infinity (i.e. null infinity $\mathscr{I}_{d+1}$).
More explicitly, the map $x^\mu\to-\frac{2x^\mu}{x^2}$ takes the simple form $u\leftrightarrow \tfrac1{r}$ in flat Bondi coordinates (see Section \ref{flatBondicords}) which makes manifest that it exchanges the two cones, i.e. $u=0\leftrightarrow r=\infty$. The d'Alembert equation is well-known to be conformally invariant, consequently, it is natural that the inversion exchanges leading modes at null infinity with leading modes on the null cone and, similarly for subleading modes. Therefore, the inversion acts as a sort of ``kinematical duality'' between the two sectors of solutions: informally speaking, it exchanges radiative modes with subradiative modes.
A more precise statement and an explanation of this relation between the two sectors can be made. 

Recall from Subsection \ref{flatBondicords} that the Minkoswki metric \eqref{metric2} in flat Bondi coordinates is related to the light-cone metric $d\tilde s^2 = du\, dv + dy^idy_i$ by a Weyl transformation $ds^2\to d\tilde s^2=\Omega^2ds^2$ with conformal factor $\Omega=\frac{|v|}{\sqrt{2}}$ . 
A conformal scalar field $\upphi$ is a conformal density of weight $w_0=-\frac{d}2$. Therefore, under a Weyl transformation it transforms as $\upphi\to\tilde\upphi=\Omega^{-\frac{d}2}\upphi$.
From this, we deduce that the wave equation on physical Minkowski spacetime with metric $ds^2$, $\Box_{{}_{\mathbb{R}^{d+1,1}}}\upphi(u,r,\mathbf y) =0$, is equivalent to the wave equation on unphysical Minkowski spacetime with conformally-related metric $d\tilde s^2$, i.e. $\widetilde\Box_{{}_{\mathbb{R}^{d+1,1}}}\tilde\upphi(u,v,\mathbf y)=0$, for the rescaled field 
\begin{equation}\label{rescaled}
\tilde\upphi(u,v,\mathbf y):=v^{-\frac{d}2}\upphi(u,v,\mathbf y)=r^{\frac{d}2}\upphi(u,r,\mathbf y)=\sum_{n=0}^{\infty} \frac{1}{r^{n}} \,\phi_n(u,\mathbf y)=\sum_{n=0}^{\infty} u^{n} \varphi_n(r,\mathbf y)\,.
\end{equation}
The latter equation reads explicitly:
\begin{equation}\label{dAlembt}
    \big(4\,\partial_u\partial_v+\partial_i\partial^i\big)\tilde\upphi(u,v,\mathbf y)=0\,.
\end{equation}
This equation, together with \eqref{rescaled}, provide another way to derive \eqref{systemd/2} and \eqref{systeqs}.
A trivial, but crucial, observation is that the Weyl transformation maps the asymptotic characteristic problem
 of solving d'Alembert equation with boundary data on the null cones $u=0$ and $r=\infty$ 
\begin{equation}
    \Box_{{}_{\mathbb{R}^{d+1,1}}}\upphi(u,r,\mathbf y) =0\,,\quad \lim_{u \to 0} \Big[r^{\frac{d}{2}}\upphi(u, r,\mathbf y)\Big]=\varphi_0(r, \mathbf y)\,,\quad \lim_{r \to \infty} \Big[r^{\frac{d}{2}} \upphi(u, r,\mathbf y)\Big]=\phi_0(u, \mathbf y)\,,
\end{equation}
to the following higher-dimensional generalisation\footnote{See e.g. the mathematical textbooks \cite{Courant} for the standard (i.e. two-dimensional) Goursat problem. For some discussions  in the physics literature about the higher-dimensionsional Goursat problem, see e.g. \cite{Neville:1971zk,Barnich:2024aln} and refs therein.} of the Goursat problem (aka ``characteristic initial value problem'') with boundary data on the null hyperplanes $u=0$ and $v=0$:
\begin{equation}\label{Goursatlightcone}
\widetilde\Box_{{}_{\mathbb{R}^{d+1,1}}}\tilde\upphi(u,v,\mathbf y)=0\,,\quad \lim_{u \to 0}\tilde\upphi(u,v,\mathbf y)=\varphi_0(v,\mathbf y)\,,\quad \lim_{v \to 0}\tilde\upphi(u, r,\mathbf y)=\phi_0(u, \mathbf y)\,,
\end{equation}
and the compatibility condition \eqref{subspaceasympt3} is mapped to the continuity condition at the intersection $u=v=0$ of the two hyperplanes: $\lim\limits_{u \to 0}\phi_0(u,\mathbf y)=\lim\limits_{v \to 0}\varphi_0(v,\mathbf y)=f(\mathbf y)$. For the well-posedness of the latter Goursat problem for smooth solutions $\tilde\upphi(u,v,\mathbf y)$, see e.g. \cite[Theorem 1]{Rendall}.

For massless fields, it is a standard trick (called the ``conformal method'' in the mathematical relativity community) to make use of a Weyl transformation to derive peeling theorems (as well as existence and uniqueness theorems) for solutions of the Cauchy problem.
In particular, one usually maps Minkowski spacetime $\mathbb{R}^{d+1,1}$ to an open subset of the Einstein universe $S^{d+1}\times \mathbb{R}$. A particular simplification comes from the fact that the conformal boundary of the ``physical'' spacetime is mapped to a mere submanifold inside the ``unphysical'' spacetime.
In this way, radiative asymptotic behaviour near infinity of Minkowski spacetime is derived from regularity conditions on the (rescaled) field in the Einstein universe (see e.g. \cite[Section 2]{Nicolas:2015lna} for a pedagogical review).
In some sense, the Weyl transformation from the metric in flat Bondi coordinates to the light-cone metric is an analogue of this standard trick, with the important difference that the present trick is well adapted to the Goursat problem while the conformal method is usually applied to the Cauchy problem (see, however, \cite{Mason:2004lqj}).

The kinetic operator in \eqref{dAlembt} is manifestly invariant under the reflection $u\leftrightarrow v$. Therefore, any solution $\tilde\upphi(u,v,\mathbf y)$ is mapped to another solution $\tilde\upphi(v,u,\mathbf y)$ via the reflection symmetry of \eqref{dAlembt}. 
In terms of the original problem, it becomes the inversion duality $u\leftrightarrow \tfrac1{r}$ which exchanges the coefficients in the two expansions: $\phi_n\leftrightarrow \varphi_n$. Consequently, we can state the following duality between modes: \textit{the inversion duality exchanges 
the leading modes at null infinity with the leading modes on the null cone}, $\phi_0\leftrightarrow \varphi_0$, and \textit{exchanges 
the subleading modes at null infinity with the subleading modes on the null cone}, $\psi_0\leftrightarrow \chi_0$. A corollary is that  \textit{the inversion duality exchanges 
the subradiative modes at null infinity with the evanescent modes on the null cone}, $\phi_0=0\leftrightarrow \varphi_0=0$.
Let us note that the inversion duality leaves the zero mode $f$ invariant (since, geometrically, the inversion preserves the cut $\mathscr{N}_{d+1}^+\cap\mathscr{I}_{d+1}^+\cong S^d$).

The inversion duality also provides an explanation for our earlier findings that special conformal transformations act on the leading modes on the null cone in the same way that translations act on the leading modes at null infinity, i.e. as specific types of shifts in $v$ and $u$, respectively. In fact, remember that translations are conjugated to special conformal transformations via an inversion, i.e. $\hat{I}\circ \hat{T} \circ \hat{I} = \hat{K}$ where $\hat{I}$ is an inversion map, while $\hat{T}$ and $\hat{K}$ are the generators of translation and special conformal transformation, respectively. In fact, the exotic Poincar\'e group $\widetilde{ISO}_0(d+1,1)$ is conjugated, via the inversion $x^\mu\to-\frac{2x^\mu}{x^2}$, to the ususal Poincar\'e group $ISO_0(d+1,1)$. 
For this reason, the vector space of solutions to \eqref{subspaceasympt2} will be denoted $\mathcal{V}^{\widetilde{\mathfrak{iso}}(d+1,1)}(\tfrac{d}{2},0)$ while the vector subspace of evanescent solutions will be denoted $\mathcal{V}^{\widetilde{\mathfrak{iso}}(d+1,1)}(\tfrac{d}{2}+1,0)$.
Accordingly, the unitary irreducible representation of the exotic Poincar\'e group $\widetilde{ISO}_0(d+1,1)$ can be denoted as
\begin{equation}
\mathcal{D}^{\widetilde{\mathfrak{iso}}(d+1,1)}(\tfrac{d}{2},0)\,:=\,\mathcal{V}^{\widetilde{\mathfrak{iso}}(d+1,1)}(\tfrac{d}{2},0)\,/\,\mathcal{V}^{\widetilde{\mathfrak{iso}}(d+1,1)}(\tfrac{d}{2}+1,0)\,\cong\, C^\infty(\mathcal{N}_{d+1})\,.
\end{equation} 
The inversion duality provides a bijective intertwiner showing the equivalence $\mathcal{D}^{\widetilde{\mathfrak{iso}}(d+1,1)}(\tfrac{d}{2},0)\cong \mathcal{D}^{\mathfrak{iso}(d+1,1)}(\tfrac{d}{2},0)$ of the two unitary irreducible representations of Poincar\'e group, in agreement with Wigner's classification: there is only one inequivalent massless scalar representation of Poincar\'e group.

\subsection{Reconstruction formula: from boundary to bulk}\label{reconstructionform}

A property that was instrumental in the seminal works relating asymptotic symmetries to soft theorems (see \cite{Strominger:2017zoo} and refs therein)
was the relation between the asymptotic behaviour of massless fields in round Bondi coordinates to (the Fourier transform over energy of) the massless field in momentum space representation. In fact, in the large radius limit a stationary phase approximation allows to show that spatial position and momenta are in some sense collinear, cf. \cite{He:2014laa} (see also \cite[Section 9, Exercise 4]{Strominger:2017zoo}).\footnote{See also \cite[Section 2.4]{Bekaert:2022ipg} for the generalisation to any spacetime dimension.} 
We will now witness the effectiveness of flat Bondi coordinates in action, that will allow to generalise this stationary phase approximation to an exact formula. 

The d'Alembert equation $\Box\upphi(x)=0$ reads in Fourier space as $p^2\upphi(p)=0$. The latter can be solved as $\upphi(p)=\delta(p^2)\phi(p)$.
We start from the Fourier transform of the on-shell scalar field
\begin{equation}\label{statph}
\upphi(u,r,\mathbf y) =\upphi(x)= \int d^{d+2}p\,\delta(p^2)\,e^{i\,p_\mu x^\mu}\phi(p)   =\int_{-\infty}^\infty d\omega \,\omega^{d-1} \int_{\mathbb{R}^d} d^d\mathbf z \,e^{-\,i\,\omega\left(u \,+\, r\,\abs{\,\mathbf{z} - \mathbf{y}}^2\right)} \,\phi(\omega, \mathbf z) 
\end{equation}
where $x^{\mu} = u\:n^{\mu} + r\:q^{\mu}(\mathbf y)$ is the position vector. In the last equality, we reduced the integral over momenta to an integral over the lightcone in momentum space, where the one-shell momentum is parametrised as $p^{\mu} = \omega\:q^{\mu}(\mathbf z)$ since it is null. We used \eqref{scalarprodq} to compute $p_\mu x^\mu$ in the exponential. For $r\gg 1$, the integrand in \eqref{statph} oscillates and the dominant contribution to the integral will be for $\mathbf{z}\approx\mathbf{y}$. In this way, we recover that indeed the vectors on the celestial plane agree in position and momentum space. But there is more. The integral over the celestial plane $\mathbb{R}^d$ on the right-hand side of \eqref{statph} is a Gaussian and can be performed exactly. Let us introduce $\mathbf{w}:=\mathbf{z} - \mathbf{y}$. Then \eqref{statph} becomes
\begin{eqnarray}
\upphi(u,r,\mathbf y) 
&=&\int_{-\infty}^\infty d\omega \,\omega^{d-1} \int_{\mathbb{R}^d} d^d\mathbf w \,e^{-\,i\,\omega\left(u + r\abs{\mathbf{w}}^2\right)} \,\phi(\omega, \mathbf y+\mathbf w)  \nonumber\\
&=&\int_{-\infty}^\infty d\omega \,\omega^{d-1} \int_{\mathbb{R}^d} d^d\mathbf w \,e^{-\,i\,\omega\left(u + r\abs{\mathbf{w}}^2\right)} \,e^{w^i\partial_i}\phi(\omega, \mathbf y)\label{intGauss}
\end{eqnarray}
It is clear that nasty things may happen at $\omega=0$. We will start by ignoring this subtlety and will return to this point later on. 

For $\omega\neq 0$, one can complete the square and do the Gaussian integral over $w^i$ in \eqref{intGauss}. The end result will soon be shown to coincide with the radiation field. We anticipate this result by writing the following equality
\begin{equation}
\label{beautiful}
    \upphi_{\text{rad}}(u,r,\mathbf y) = \left(\frac{\pi}{i}\right)^{\frac{d}{2}} \frac{1}{r^{\frac{d}{2}}} \int_{-\infty}^\infty  d\omega\,\omega^{\frac{d}{2}-1}\, e^{-\,i\,\left(\omega u+\frac{\partial_i\partial^i}{4\omega r}\right)}  \phi(\omega, \mathbf y) 
\end{equation}
This reconstruction formula gives us an exact expression (to all orders in $1/r$) of the radiative field $\upphi_{\text{rad}}(u,r,\mathbf y)$ in terms of the field $\phi(\omega, \mathbf y)$ defined on the light-cone in momentum space. To check this identification \eqref{beautiful}, let us start by considering the radiative data
\begin{equation}
\label{mod_fourier_1}
 \phi_0(u,\mathbf y) =    \lim_{r \to \infty} \big[\,r^{\frac{d}{2}}\upphi_{\text{rad}}(u,r,\mathbf y)\,\big] = \left(\frac{\pi}{i}\right)^{\frac{d}{2}}  \int_{-\infty}^\infty  d\omega \,\omega^{\frac{d}{2}-1} \,e^{-\,i\,\omega u}  \phi(\omega, \mathbf y)\,. 
\end{equation}
Thus, we recover that the radiative data $\phi_0(u,\mathbf y)$ is a weighted Fourier transform over the energy $\omega$ of the field $\phi(\omega, \mathbf y)$ in momentum representation (compare with \cite[Eq.(50)]{Bekaert:2022ipg} in round Bondi coordinates). The relation \eqref{mod_fourier_1} can be used to check the property $\langle\upphi \mid \upphi'\rangle=\langle\,\phi_0 \mid \phi'_0\,\rangle$ mentioned towards the end of Section \ref{expnullinfty}. If we further take the limit $u\to 0$ in \eqref{mod_fourier_1}, then we get the boundary data \eqref{subspaceasympt3} at the cut $(u,r,\mathbf y)=(0,\infty,\mathbf y)$:
\begin{equation}
\label{mod_fourier_1'}
f(\mathbf y)=    \lim_{u \to 0}\phi_0(u,\mathbf y) =  \left(\frac{\pi}{i}\right)^{\frac{d}{2}}  \int_{-\infty}^\infty   d\omega \,\omega^{\frac{d}{2}-1}\,\phi(\omega, \mathbf y)\,. 
\end{equation}
In other words, the radiative data restricted to the good cut corresponding to the origin is proportional to the Mellin transform\footnote{A minor thing to note is integral in the Mellin transforms goes from $0$ to $\infty$, but the integral here takes the limits from $-\infty$ to $+\infty$, but we can still call it Mellin transforms by taking even/odd part of the function $\phi(\omega,\mathbf y)$.} over the energy $\omega$, evaluated at $\tfrac{d}{2}$. 

Now the idea is to interpret $1/\omega$ as the Fourier transform of the formal operator $i(\partial_u)^{-1}$. 
Of course, this requires some care and is somewhat ambiguous (since $1/\omega$ is not defined at $\omega=0$ and, equivalently, the partial derivative $\partial_u$ has a kernel). The prescription that implements the identification \eqref{beautiful} is to replace each $1/\omega$ with the operator defined by $i(\partial_u)^{-1}g(u)\,:=\,i\int_0^u du'g(u')$ for any smooth function of retarded time (i.e. we set to zero all ``integration constants''). 
In this way, one can write
\begin{equation}
    \upphi_{\text{rad}}(u,r,\mathbf y) = \left(\frac{\pi}{i}\right)^{\frac{d}{2}} \frac{1}{r^{\frac{d}{2}}} \int_{-\infty}^\infty  d\omega\,\omega^{\frac{d}{2}-1}\, e^{-\frac{\partial_i\partial^i}{4r}(\partial_u)^{-1}}\,e^{-\,i\,\omega u} \, \phi(\omega, \mathbf y) 
\end{equation}
Using \eqref{mod_fourier_1} gives the explicit expression of the radiation field
\begin{equation}\label{explicitrad}
    \upphi_{\text{rad}}(u,r,\mathbf y) = \frac{e^{-\frac{\partial_i\partial^i}{4r}(\partial_u)^{-1}}}{r^{\frac{d}{2}}}\,\,\phi_0(u, \mathbf y) \,,
\end{equation}
which agrees with \eqref{leadingpiece}-\eqref{tail}. This proves our claim.

Endowed with the reconstruction formula \eqref{beautiful} of the radiation field in position space in terms of the field in momentum space, we can also take a different perspective. Let us assume \eqref{mod_fourier_1} is invertible, in the sense that 
\begin{equation}\label{inverseF}
\phi(\omega, \mathbf y) = \left(\frac{i}{\pi}\right)^{\frac{d}{2}} \frac{1}{2\pi \omega^{\frac{d}{2}-1}} \int_\infty^\infty du'\,e^{i\omega u'}\,\phi_0(u',\mathbf y) 
\end{equation}  
This formula holds, for instance, if $\omega^{\frac{d}{2}-1}\,\phi(\omega, \mathbf y)$ is a tempered distribution (or, equivalently, if $\phi_0(u,\mathbf y)$ is a tempered distribution (see e.g. \cite[Section 6.3]{Kanwal}). 
Plugging \eqref{inverseF} into \eqref{beautiful} yields 
\begin{equation}
\upphi_{\text{rad}}(u,r,\mathbf y)  
=\frac{1}{2 \pi r^{\frac{d}{2}}} \int_{-\infty}^\infty du' \left[\, \int_{-\infty}^\infty d\omega \,e^{-i\omega( u - u')}\, e^{-i \frac{\partial_i\partial^i}{4\omega r}} \,\right]  \phi_0(u',\mathbf y)    
\end{equation}
It turns out to be convenient to go into the Fourier space of the celestial plane $\mathbb{R}^d$, with variables $\mathbf k$ dual to $\mathbf y$, and to make use of the rescaled energy $\omega':=\frac{4r}{|\mathbf k|^2}\,\omega$. In this way, one ends up with the following \textit{boundary-to-bulk} reconstruction formula:
\begin{equation}\label{bdytobulk}
\upphi_{\text{rad}}(u,r,\mathbf k) 
= \frac{|\mathbf k|^2}{4r^{\frac{d}{2}+1}} \int_{-\infty}^\infty du' \,
\,K\left(\frac{|\mathbf k|^2}{4r}\,(u-u')\right)
\,\phi_0(u',\mathbf k)    
\end{equation}
where the generalised function of one variable
\begin{equation}\label{K(uz)}
K(y)\, := \frac{1}{2\pi} \int_{-\infty}^\infty d\omega \,e^{-i\,\omega y}\, e^{+\frac{i}{\omega}}= \, \delta(y) + \frac{1}{2}\, sgn(y) \frac{d}{dy}\Big(J_0(2i\sqrt{y})\Big)
\end{equation}
can be interpreted as a sort of \textit{boundary-to-bulk propagator}, with $J_0$ the Bessel function of the first kind.
The distribution \eqref{K(uz)} is nothing but the inverse Fourier transform of the function $e^{\frac{i\,}{\omega}}$. Its explicit computation can be found in Appendix \ref{Kprop}. 
Note that the Dirac delta in \eqref{K(uz)} is the term necessary for reproducing the radiative data. In fact, 
\begin{eqnarray}
\upphi_{\text{rad}}(u,r,\mathbf k) 
&=& \frac{|\mathbf k|^2}{4r^{\frac{d}{2}+1}} \int_{-\infty}^\infty du' \,
\,\Big[\,\delta\left(\frac{|\mathbf k|^2}{4r}\,(u-u')\right)+\text{regular}
\Big]\,\phi_0(u',\mathbf k) \\
&=& \frac{1}{r^{\frac{d}{2}}} \Big(\int_{-\infty}^\infty du' \,
\,\delta(u-u')\phi_0(u',\mathbf k)\,+\,\phi_{\text{tail}}(u,r,\mathbf k)  \Big)\,,
\end{eqnarray}
where the notation $\phi_{\text{tail}}$ has been introduced for the integrals involving the  regular  piece, in order to underline that the end result has the same form as \eqref{leadingpiece}.
In other words, the nontrivial part of the boundary-to-bulk formula \eqref{bdytobulk} is the one for the tail
\begin{equation}\label{bdytobulk2}
\phi_{\text{tail}}(u,r,\mathbf k) 
= \frac{|\mathbf{k}|^2}{4r}\int_{-\infty}^\infty du' \,\,K_{\text{reg}}\left(\frac{|\mathbf k|^2}{4r}\,(u-u')\right)
\,\phi_0(u',\mathbf k)    
\end{equation}
where 
\begin{equation}\label{Kreg}
K_{\text{reg}}(y)\, :=  \frac{1}{2}\, sgn(y) \frac{d}{dy}\Big(J_0(2i\sqrt{y})\Big)=\frac{1}{2}\, sgn(y)\sum_{n=0}^{\infty} \frac{y^{n}}{(n+1)!\,n!}\,.
\end{equation}

Furthermore, we can also put $u = 0$ in the boundary-to-bulk formula \eqref{bdytobulk} and get
\begin{equation}
\varphi_0(r,\mathbf k) =   \frac{|\mathbf{k}|^2}{4r}\int_{-\infty}^\infty du'  K\left(-\frac{|\mathbf k|^2}{4r}\,u'\right) \phi_0(u',\mathbf k)
\end{equation}
This can be interpreted as a sort of scattering formula, where we put in radiative data at null infinity $\phi_0$ and evolve (or, rather, de-evolve) it to give the value $\varphi_0$ of the radiation field on the null cone through the origin.

As mentioned above, nasty things may happen at the tip of the light-cone, $\omega=0$, at various steps in the previous chain of arguments. Ignoring related functional subtleties typically resulted in missing the subradiative (aka ``chthonian'') sector which, in the momentum representation, is somewhat hidden (in the ``underworld'' below the visible radiation field). For instance, such subtleties showed up in the ambiguity appearing when one wants to reinterpret $1/\omega$ as the Fourier transform of the operator $i(\partial_u)^{-1}$. Fixing this ambiguity in a suitable way lead to the radiation field. However, by taking this ambiguity at face value, one should be able to recover the subradiative modes. We do not aim at full mathematical rigor on this tricky point that would deserve further investigation. Instead,we  present several heuristic arguments in favor of a full reconstruction formulae (involving both sectors) in Appendix \ref{heuristic}. 

\section{Massless spin one field}\label{Max}

It has been known for a very long time \cite{Bateman} that conformal transformations are
the most general coordinate transformations preserving the form of Maxwell equations on four-dimensional Minkowski spacetime. It is also well-known that Maxwell equations, minimally coupled to a curved background, are invariant under Weyl transformations. In Subsection \ref{invMax}, we write Maxwell equations in the light-cone-like coordinates $(u,v,y^i)$ in order to see explicitly that, in spacetime dimension four, the resulting equations take the form of Maxwell equations in the background lightcone metric. In Subsection \ref{Maxasympt}, we show explicitly that the inversion symmetry indeed exchanges again the radiative and subradiative sectors.

\subsection{Inversion symmetry}\label{invMax}

In flat Bondi coordinates, an important corollary of the celebrated result on Weyl invariance is the following: the solutions of Maxwell equations on four-dimensional Minkowski spacetime expressed in the flat Bondi coordinates, adapted for the asymptotic expansions in $v=1/r$ and/or in $u$, can be mapped to solutions of Maxwell equations in light-cone coordinates $u$ and $v$. In this way, the inversion duality between modes at null infinity and on null cone is merely the reflection symmetry $u\leftrightarrow v$. 
In Section \ref{anyspin}, this will be generalised to any integer spin in any even dimension by making use of generalised spin-$s$ Bargmann-Wigner equations (remember that the spin-1 Bargmann-Wigner equations are just Maxwell equations). We spell out the spin-one case in many details because it provides a simple example where all computations can be made entirely explicit without effort.

The Maxwell equations read, in an arbitrary metric $g_{\mu\nu}$, as
\begin{equation}\label{Maxwell}
\nabla_{[\mu} F_{\nu \rho]} = 0\,, \qquad\nabla^\mu F_{\mu \nu} = 0\,, 
\end{equation}
in terms of the Faraday tensor (aka electromagnetic field-strength) $F_{\mu\nu}$ and of the Levi-Civita connection $\nabla$.
Let us write the explicit form of the system \eqref{Maxwell} in the metric \eqref{metric2} in terms of the coordinates $(u,v,y^i)$ on Minkowski spacetime $\mathbb{R}^{d+1,1}$. The first equation in \eqref{Maxwell} gives the following equations:
\begin{equation}\label{Max1}
\partial_{u}F_{vi} + \partial_{v}F_{iu} + \partial_{i}F_{uv} = 0\,, \qquad \partial_{i}F_{jk} + \partial_{k}F_{ij} + \partial_{j}F_{ki} = 0\,,
\end{equation}
\begin{equation}\label{Max2}
\partial_{i}F_{ju} + \partial_{u}F_{ij} + \partial_{j}F_{ui} = 0\,, \qquad \partial_{i}F_{jv} + \partial_{v}F_{ij} + \partial_{j}F_{vi} = 0\,.
\end{equation}
These equations are manifestly symmetric under the exchange $u \leftrightarrow v$. More precisely, the two equations in \eqref{Max1} are invariant by themselves, while the transformation $u \leftrightarrow v$ exchanges the two equations in \eqref{Max2}.
Now consider the second equation in \eqref{Maxwell} for the various possible values of the index $\nu$:
\begin{equation}
\label{eq:barg_wig1}
\nu = v : \quad \partial_v F_{uv} + \frac{1}{2} \partial^{i} F_{iv} + \frac{2-d}{v} F_{uv} = 0 \quad\Longleftrightarrow\quad \partial_v\left( v^{-d+2} F_{vu}\right) = \frac{1}{2} \partial^{i}\left(v^{-d+2} F_{iv}\right)\,,
\end{equation}
\begin{equation}
\label{eq:barg_wig2}
\nu = u : \quad \partial_u F_{uv} = \frac{1}{2}\,\partial^i F_{iu}\,,
\end{equation}
\begin{eqnarray}
\label{eq:barg_wig3}
&\nu = j :& \quad \partial_u F_{vj} + \partial_v F_{uj} + \frac{1}{2}\partial^i F_{ij} + \frac{2-d}{v} F_{uj} = 0 \nonumber\\&&\quad\Longleftrightarrow\quad\partial_v \left(v^{-d + 2} F_{uj} \right) + \partial_u \left(v^{-d + 2} F_{vj} \right) + \frac{1}{2} \partial^i\left(v^{-d + 2} F_{ij} \right) = 0\,.
\end{eqnarray}
We see that, when $d = 2$ (i.e. in $D=4$ space-time dimensions), the equation (\ref{eq:barg_wig1}) goes to (\ref{eq:barg_wig2}) (and vice-versa) under the exchange $u \leftrightarrow v$, while the equation and (\ref{eq:barg_wig3}) is invariant by itself.
But there is more. When $d = 2$, the equations \eqref{Max1}-\eqref{eq:barg_wig3} are nothing but Maxwell equations  in the light-cone metric $d\tilde s^2 = du\, dv + dy^idy_i$ on $\mathbb{R}^{3,1}$ (therefore, it is obvious that they are invariant under the reflection $u\leftrightarrow v$).
This result could have been obtained without performing any computation since one knows that Maxwell equations on any (possibly curved) four-dimensional spacetime are invariant under Weyl transformations ($ds^2\to d\tilde s^2=\Omega^2ds^2$) of the background metric, where the Faraday tensor $F_{\mu\nu}$ does not transform under Weyl transformations. 
This observation is useful because it allows to generalise these results to all integer spin and all even dimensions for wave equations which have Weyl (and, hence, conformal) symmetry around a conformally-flat background. A subtlety for $s\neq 1$ is that the field-strength has non-vanishing conformal weight so the original Bargmann-Wigner equations on Minkowski spacetime in flat Bondi coordinates $(u,r,y^i)$ are not mapped exactly to Bargmann-Wigner equations in the lightcone coordinates $(u,v,y^i)$, because the field-strengths there come accompanied with factors which are powers of $v$.

\subsection{Asymptotic expansion}\label{Maxasympt}

Let us now assume that the field-strength admits a Laurent series expansion in $v$,
\begin{equation}\label{Laurent}
    F_{\mu\nu}(u,v,\mathbf y)  = \sum\limits_{n\,=\,\bar n}^\infty v^n F^{(n)}_{\mu\nu}(u,\mathbf y) \,,
\end{equation}
where $\bar n\in\mathbb{Z}$ is an integer.
The system \eqref{Max1}-\eqref{eq:barg_wig3} then becomes (the seven equations are ordered differently),
\begin{eqnarray}
 \quad & \partial{}^{}_{[i}F_{jk]}^{(n)} = 0\,,\label{max_7}\\
\quad & \dot{F}_{ij}^{(n)} + \partial{}^{}_{[i} F_{j]u}^{(n)} = 0 \,,\label{max_5}\\
 \quad & \dot{F}_{uv}^{(n)} = \frac{1}{2} \partial^i F_{iu}^{(n)} \,,\label{max_6}\\
 \quad & \dot{F}_{v i}^{(n-1)} + \partial_i F_{u v}^{(n-1)} - n F_{ui}^{(n)} = 0\,, \label{max_3}\\
  \quad & \dot{F}_{vi}^{(n-1)} +  (2 - d + n) F_{ui}^{(n)} + \frac{1}{2} \partial^j F_{ji}^{(n-1)} = 0 \,,\label{max_4}\\
 \quad & n F_{ij}^{(n)} + \partial{}^{}_{[i} F_{j]v}^{(n-1)} = 0\,, \label{max_1}\\
 \quad & (2 - d + n) F_{uv}^{(n)} + \frac{1}{2} \partial^i F_{iv}^{(n-1)}=0\,, \label{max_2}
\end{eqnarray}
where the square bracket denotes antisymmetrisation (e.g. $T_{[ij]}:=T_{ij}-T_{ji}$) and the dot is a shorthand for the derivative $\partial_u$ with respect to retarded time. The first equation \eqref{max_7} merely states that each spatial two-form $F_{ij}^{(n)}$ is closed, as will be implicitly assumed from now on. We show in Appendix \ref{Maxapp} that the second and third equations, \eqref{max_5} and \eqref{max_6}, are redundant when $n \neq 0$ and when $n \neq d - 2$ respectively.
Adding the fourth and fifth equations, $\eqref{max_3}$ and $\eqref{max_4}$, gives
\begin{equation}
    2\dot{F}_{v i}^{(n-1)} + \partial_i F_{u v}^{(n-1)}  + (2 - d) F_{ui}^{(n)}  + \frac{1}{2} \partial^j F_{ji}^{(n-1)} = 0\,. \label{max_add}
\end{equation}
For $d\neq 2$, this allows to express $F_{ui}^{(n)}$ in terms of lower-order components $F_{\mu\nu}^{(n-1)}$.
For $d=2$, the equation \eqref{max_add} becomes a homogenous equation (i.e. involving coefficients at the same order $n$)
\begin{equation}
    \dot{F}_{iv}^{(n)} = \frac12\,\partial_i F_{u v}^{(n)}   + \frac{1}{4} \partial^j F_{ji}^{(n)}\,.  \label{max_add4d}
\end{equation}
Subtracting the fourth equation, \eqref{max_3}, from the fifth equation, \eqref{max_4}, gives
\begin{equation}
     (2 - d + 2n) F_{ui}^{(n)} - \partial_i F_{u v}^{(n-1)} + \frac{1}{2} \partial^j F_{ji}^{(n-1)} = 0\,. \label{max_sub}
\end{equation}
For $2 - d + 2n\neq 0$, this also allows to express $F_{ui}^{(n)}$ in terms of lower-order components $F_{\mu\nu}^{(n-1)}$.
Finally, the sixth equation \eqref{max_1} for $n\neq 0$ and the seventh equation \eqref{max_2} for $2-d+n\neq 0$ allow to express respectively $F_{ij}^{(n)}$ and $F_{uv}^{(n)}$ in terms of some lower-order components $F_{\mu\nu}^{(n-1)}$.

We set $d = 2$ from now on (in this section). For $n=0$, the equations \eqref{max_5}-\eqref{max_6} are relevant and read
\begin{equation}
     \dot{F}_{ij}^{(0)} =-\partial{}^{}_{[i} F_{j]u}^{(0)} \,, \qquad \dot{F}_{uv}^{(0)} = \frac{1}{2} \partial^i F_{iu}^{(0)}
\end{equation}
which can be solved  explicitly as 
\begin{equation}
    F_{ij}^{(0)}(u,\mathbf y)  = -\int_0^u du'\,\partial{}^{}_{[i} F_{j]u}^{(0)}(u',\mathbf y) \, +\, \tilde{F}_{ij}^{(0)}(\mathbf y)\,, \quad {F}_{uv}^{(0)}(u,\mathbf y) = \frac{1}{2} \int_0^u  du'\,\partial^i F_{iu}^{(0)} (u',\mathbf y) \,+\, \tilde{F}_{uv}^{(0)}(\mathbf y)\,, \label{uv_iu}
\end{equation}
where $\tilde{F}^{(0)}_{ij}(\mathbf y)$ is a closed two-form on the plane and $\tilde{F}^{(0)}_{uv}(\mathbf y)$ is an arbitrary function on the plane. 
Taking into account \eqref{uv_iu}, the equation \eqref{max_add4d} for $n=0$ can be solved as
\begin{eqnarray}
      F_{iv}^{(0)}(u,\mathbf y)& = & \frac{1}{4}\int_0^udu' \int_0^{u'} du''\, \partial_i\partial^j F_{ju}^{(0)}(u'',\mathbf y) \, +\, \frac{u}2 \,\partial_i\tilde{F}_{uv}^{(0)}(\mathbf y)\nonumber\\&&\quad -\,\frac{1}{4}\int_0^u du' \int_0^{u'} du''\, \partial^j \partial{}^{}_{[j} F_{i]u}^{(0)}(u'',\mathbf y) \, +\, \frac{u}{4}\,\partial^j\tilde{F}_{ji}^{(0)}(\mathbf y) \,+\,  \tilde{F}_{iv}^{(0)}(\mathbf y)\,,\label{Fiv0}
\end{eqnarray}
where $\tilde{F}_{iv}^{(0)}(\mathbf y)$ is an arbitrary function on the plane. As one can see, all coefficients $F_{\mu\nu}^{(0)}$ at order zero are determined in terms of;
\begin{itemize}
    \item a one-form $F_{iu}^{(0)}(u,\mathbf y)=F_{iu}(u,v=0,\mathbf y)$ at null infinity $\mathscr{I}_3$, which will be called the \textit{radiative data},
    \item and a collection of forms on the plane $\mathbb{R}^2$ with all possible degrees: a closed two-form $\tilde{F}^{(0)}_{ij}(\mathbf y)$, an arbitrary one-form $\tilde{F}_{iv}^{(0)}(\mathbf y)$ and a zero-form (i.e. a function) $\tilde{F}_{uv}^{(0)}(\mathbf y)$.
\end{itemize}

For $d=2$ and $n=0$, the remaining equations
\eqref{max_3}-\eqref{max_2} impose differential conditions on the coefficients $F_{\mu v}^{(-1)}$. More precisely, $F_{u v}^{(-1)}$ does not depend on $y^i$ (i.e. $\partial_i F_{u v}^{(-1)}=0$) and the one-form  $F_{iv}^{(-1)}$ on the celestial plane $\mathbb{R}^2$ is harmonic (i.e. $\partial{}^{}_{[i} F_{j]v}^{(-1)} = 0$ and $\partial^i F_{iv}^{(-1)}=0$).
Since the latter one-form should have a smooth extension to the whole celestial sphere $S^2$, the harmonicity actually implies that it must vanish (since the first Betti number of the sphere vanishes): $F_{i v}^{(-1)}=0$. 
In turn, the equations \eqref{max_7} for $n=-1$ and \eqref{max_4} for $n=0$ imply that the two-form $F_{ij}^{(-1)}=\varepsilon_{ij} f^{(-1)}$ is also harmonic (i.e. closed and coclosed). 
Therefore $f^{(-1)}=\frac12\varepsilon^{ij}F_{ij}^{(-1)}$ does not depend on $y^i$. 
These stringent conditions motivate us to focus on solutions with  $F_{\mu\nu}^{(-1)}=0$ for all components. 

From now on we will  focus on the solutions of Maxwell equations which are of the form \eqref{Laurent} with $\bar n=0$ or, in other words, to solutions which have a (possibly formal) power series expansion in the coordinate $v$, i.e. $F_{\mu\nu}(u,v,\mathbf y)  = \sum\limits_{n=0}^\infty v^n F^{(n)}_{\mu\nu}(u,\mathbf y)$. We will call them \textit{radiative solutions} to Maxwell equations (in spacetime dimension four). Note that in the flat Bondi coordinates, this translates into the asympototic expansions of the form
\begin{equation}\label{Laurent1}
    F_{ij}(u,v,\mathbf y)  = \sum\limits_{n=0}^\infty \frac1{r^n} F^{(n)}_{ij}(u,\mathbf y)\,,\qquad F_{iu}(u,v,\mathbf y)  = \sum\limits_{n=0}^\infty\frac1{r^n}F^{(n)}_{iu}(u,\mathbf y)\,,
\end{equation}
\begin{equation}\label{Laurent2}
 F_{ir}(u,v,\mathbf y)  = \frac1{r^2}\sum\limits_{n=0}^\infty \frac1{r^n} F^{(n)}_{ir}(u,\mathbf y)\,,\qquad F_{ur}(u,v,\mathbf y)  = \frac1{r^2}\sum\limits_{n=0}^\infty \frac1{r^n} F^{(n)}_{ur}(u,\mathbf y) \,.
\end{equation}
This behaviour is in agreement with the usual behaviour in round Bondi coordinates, see e.g. Equation (2.34) in \cite{Godazgar:2021iae}.
Finally, we will call \textit{subradiative solutions}, the radiative solutions such that $F^{(0)}_{\mu\nu}(u,\mathbf y)=0$ or, in other words, $F_{\mu\nu}(u,v,\mathbf y)  = \sum\limits_{n=1}^\infty v^n F^{(n)}_{\mu\nu}(u,\mathbf y)$.

For $d=2$ and $n\geqslant 1$, the three equations \eqref{max_1}, \eqref{max_2} and \eqref{max_sub} allow to express respectively $F_{ij}^{(n)}$, $F_{uv}^{(n)}$ and $F_{ui}^{(n)}$ in terms of lower-order components, while the equation \eqref{max_add4d} which is first-order in $v$ can be solved
to express ${F}_{iv}^{(n)}$ in terms of lower components, up to ``integrating constants'' $\tilde{F}_{iv}^{(n)}(\mathbf{y})$. The infinite tower of arbitrary one-forms $\tilde{F}_{iv}^{(n)}$ on the plane $\mathbb{R}^2$, with $n\geqslant 0$, can be packaged into a single generating function,
\begin{equation}
\sum\limits_{n\,=\,0}^\infty v^n\tilde{F}^{(n)}_{iv}(\mathbf y) =
{F}_{iv}(u=0,v,\mathbf y)
\,,    
\end{equation}
which will be called the \textit{subradiative data}. It is the necessary and sufficient data for subradiative solutions.

To summarise, the boundary data that we need to specify in order to have a unique solution (up to scalar soft modes\footnote{A subtlety is indeed that, strictly speaking, we also need two more functions on the plane, $f_1(\mathbf y):=\varepsilon^{ij}\tilde{F}^{(0)}_{ij}(\mathbf y)$ and $f_2(\mathbf y):=\tilde{F}^{(0)}_{uv}(\mathbf y)$, apart from the radiative and subradiative boundary data.} which are not our main focus here) consists of two arbitrary $1$-forms: one at null infinity, the radiative data $F_{iu}(u,v=0,\mathbf y)$, and one on the interior null cone, the subradiative data $F_{iv}(u=0,v,\mathbf y)$. They have a particularly simple interpretation in the light-cone metric: they are data along the two null hyperplanes $u=0$ and $v=0$. 
To conclude our analysis of Maxwell equations, we go back to the arguments about the inversion duality. We already know that the exchange $u  \leftrightarrow v$ is a symmetry. Therefore, the subradiative data $F_{iv}(u = 0,v,\mathbf y)$ is exchanged with the radiative data $F_{iu}(u,v=0,\mathbf y)$. So the duality between radiative and subradiative modes follows here too. As argued in the next section, these features are not specific to spin $0$ and $1$: on the contrary, they are valid for any integer spin.

\section{Generalisation to any spin}\label{anyspin}

As explained in the previous sections, the inversion duality between the radiative and subradiative sectors amounts, in practice, to a precise duality between the expansions in $1/r$ and $u$ under an inversion, expressed in flat Bondi coordinates as $u\leftrightarrow 1/r$. The existence and precise form of the latter transformation of the solutions (i.e. the analogue of Sections \ref{dalembert}-\ref{Max}) will follow as a corollary of the Weyl covariance of Bargmann-Wigner equations on a conformally-flat background. 

Let us recall once again the logic of our argument. First, the Minkowski metric \eqref{metric2} is related by a Weyl transformation $ds^2\to d\tilde s^2=\Omega^2ds^2$ with conformal factor $\Omega=\frac{|v|}{\sqrt{2}}$ to the light-cone metric \eqref{lightconmetric} with $x^+=u$ and $x^-=-v/2$. Second, the Weyl covariance of Bargmann-Wigner equations can be used to map the original equations to their simple form in terms of the light-cone-like coordinates $(u,v,y^i)$. Third, the solutions of latter equations can be computed explicitly, they admit an asymptotic expansion in $u$ and $v$ and are expressed in terms of two boundary data (at $u=0$ and at $v=0$, respectively). Fourth, the action of the inversion duality $u\leftrightarrow v$ on the solutions can be computed explicitly from the transformation law of the field-strength under a conformal transformation. 

Let us return to Weyl covariance. In four dimensions, the proof of the  Weyl symmetry of Bargmann-Wigner equations for all spins dates back to Penrose's works \cite{Penrose:1962ij,Penrose:1965am} on the peeling properties of massless fields (more precisely, see Sections 10 and 12 in \cite{Penrose:1965am}). In higher dimensions, the Weyl symmetry of generalised Bargmann-Wigner equations on conformally-flat spacetime follows\footnote{In fact, let us stress that, although the ambient formulation on $\mathbb{R}^{d+2,2}$ is used to make manifest the conformal invariance of wave equations, the ambient formulation actually also ensures their Weyl covariance, since changes of conformal representative correspond to different choices of sections of the ambient null cone (aka ``metric bundle'' in Fefferman-Graham terminology \cite{Fefferman:2007rka}).} from their ambient formulation.
The generalised Bargmann-Wigner equations on flat spacetime that describe spinning ``singletons'' (i.e. unitary irreducible representations of the Poincar\'e group that lift to the restricted conformal group, see e.g. the review \cite{Bekaert:2011js} and refs therein) are known for some time \cite{Bandos:2005mb}, as well as the fact that they admit an ambient formulation \cite{Arvidsson:2006fq}. Although the existence of an ambient formulation for these equations guarantees their Weyl symmetry (hence, their conformal symmetry and, hence, their inversion symmetry), we nevertheless provide a direct proof of this fact in Subsection \ref{weylcov} because it is instructive by itself and, moreover, it requires a careful discussion of the conformal weight of field-strengths, which will be necessary for deriving the explicit expressions of the inversion duality for the various components in the general solution.

To conclude this paper, some group-theoretical comments are in order. In Subsection \ref{groupth}, we recall some facts of representation theory in order to try inserting our explicit results (in flat Bondi coordinates) into a framework more adapted for statements which are independent of the specific choice of formulation.

\subsection{Weyl covariance of generalised Bargmann-Wigner equations}\label{weylcov}

The compact version of generalised Bargmann-Wigner equations for bosonic singletons of spin $s\in\mathbb N$ on any conformally-flat spacetime of any even dimension $D=d+2$ in terms of multi-forms\footnote{For their use in the description of general mixed symmetry tensor fields, see e.g. \cite{Bekaert:2002dt}.}, i.e. fields depending on Grassmann-even coordinates $x^\mu$ (with $\mu,\nu=0,1,\ldots,D-1$) and Grassmann-odd coordinates $\theta_I^\nu$ (with $I,J=1,\ldots,s$). More precisely, these generalised Bargmann-Wigner equations are a linear system of equations (some algebraic some differential) expressed in terms of the field-strength $K(x^\mu,\theta^\nu_I)$ which is a multiform.

We define the following algebraic operators\begin{equation}
N_I^J=\theta_{{}_I}^\mu\frac{\partial}{\partial\theta_{{}_J}^\mu}\,,\qquad G_{IJ}=g_{\mu\nu}\,\theta_{{}_I}^\mu \theta_{{}_J}^\nu\,,\qquad T^{IJ}=g^{\mu\nu}\,\frac{\partial}{\partial \theta_{{}_I}^\mu}\frac{\partial}{\partial \theta_{{}_J}^\nu}\,.
\end{equation}
where $g_{\mu\nu}$ is a conformally-flat metric.
Among the generalised Bargmann-Wigner equations, the algebraic equations are the system
\begin{equation}\label{Bianchi}
(N_I^J-\tfrac{D}{2}\,\delta_I^J\,)\,K=0\,, \qquad T^{IJ}K=0\,,    
\end{equation}
which express that the on-shell field-strength carries an irreducible representation of the Lorentz algebra $\mathfrak{so}(d+1,1)$ labeled by a rectangular Young diagram of length $s$ and height $\tfrac{D}{2}=\tfrac{d}{2}+1$.
The condition on the height follows from the first equation in \eqref{Bianchi} with $I=J$, i.e. it reads $N_I^IK=\tfrac{D}{2}$ (without sum over the repeated column index $I$) that .
We also define the algebraic operator 
\begin{equation}
    N=\sum\limits_{I=1}^sN_I^I
\end{equation} for later purpose. This operator counts the total number of indices of a multiform, in particular $NK=s\,\tfrac{D}{2}\,K$\,.

The differential equations read
\begin{equation}\label{closure}
\nabla_I\,K=0\,,\quad\overline\nabla_J\,K=0\,,\quad(I,J=1,\ldots,s)
\end{equation}
where the covariant curls and co-curls are defined as
\begin{equation}
\nabla_I=\theta_{{}_I}^\mu\,\nabla_\mu\,,\qquad \overline\nabla^I=g^{\mu\nu}\nabla_\mu\,\frac{\partial}{\partial\theta_{{}_I}^\nu}\,,
\end{equation}
and $\nabla$ denotes the Levi-Civita connection of the conformally-flat metric $g_{\mu\nu}\,$.

Recall that, under a Weyl transformation
\begin{equation}\label{Weyl}
    g'_{\mu\nu}=e^{2\omega}g_{\mu\nu}\quad\Longleftrightarrow\quad g'^{\mu\nu}=e^{-2\omega}g^{\mu\nu}\,,
\end{equation}
the Levi-Civita connection transforms as
\begin{equation}
    \Gamma'^{\rho}_{\mu\nu}=\Gamma^{\rho}_{\mu\nu}+\delta^\rho_\mu\,\partial_\nu\omega+\delta^\rho_\nu\,\partial_\mu\omega
    -g_{\mu\nu}\,g^{\rho\sigma}\partial_\sigma\omega\,.
\end{equation}
Accordingly, the covariant curls and co-curls transform as
\begin{eqnarray}\label{nabla'1}
    \nabla'_I&=&\nabla_I\,+\,\sum\limits_{J=1}^s\nabla_J\,\omega\,(N_I^J\,-\delta_I^J\,N)\,+\sum\limits_{J=1}^s\,G_{IJ}\,\,\overline\nabla^J\omega\,,\\
    \overline\nabla'^I&=&\overline\nabla^I\,+\,\,\sum\limits_{J=1}^s(N_J^I\,-\delta^I_J\,N)\,\overline\nabla^J\omega\,+\,\sum\limits_{J=1}^s\nabla_J\,\omega\,\,T^{IJ}\,.\label{nabla'2}
\end{eqnarray}
Let us assume that the field-strength transforms as a conformal density of weight $w_s\in\mathbb R$, i.e. $K'=e^{w_s\,\omega}K$, in such a way that the algebraic equations \eqref{Bianchi} are obviously preserved by a Weyl transformation \eqref{Weyl}.
However, the differential equations are only preserved  if the conformal weight is suitably fixed. In fact, the transformation laws \eqref{nabla'1}-\eqref{nabla'2} of the (co)curls implies that the differential equations \eqref{closure} are mapped under a Weyl transformation to
\begin{eqnarray}\label{nabla1}
    \nabla'_I(e^{w_s\,\omega}K)&=&\nabla_I\,\omega\,(w_s\,+\,N_I^I\,-\,N+s-1)\,K=0\,,\\
    \overline\nabla'^I(e^{w_s\,\omega}K)&=&\overline\nabla^J\omega\,(w_s\,+\,N_J^I\,+\,N+s-1)\,K=0\,.\label{nabla2}
\end{eqnarray}
where there is no sum over $I$ in the right-hand side and we used the operatorial identity
\begin{equation}
\sum\limits_{J=1}^s\,[G_{IJ},\overline\nabla^J\omega]=(s-1)\nabla_I\,\omega\,.
\end{equation}
Finally, one should recall the conditions $N_I^IK=\tfrac{D}{2}$ and $NK=s\,\tfrac{D}{2}\,K$ to conclude that Weyl covariance of generalised Bargmann-Wigner equations will be satisfied for all $\omega$ iff 
\begin{equation}
\Big(w_s-\left(\tfrac{D}{2}-1\right)(s-1)\Big)\,K=0\,.
\end{equation}
In conclusion, the conformal weight of the field-strength is fixed and 
\begin{equation}
    K'=e^{w_s\,\omega}K\,,\quad\text{with}\quad w_s=\left(\tfrac{D}{2}-1\right)(s-1)=\tfrac{d}{2}\,(s-1)\,,
\end{equation}
under a Weyl transformation \eqref{Weyl}. For $s=0$, we recover that $w_0=\tfrac{d}{2}$ for the conformal scalar field $\upphi$. For $s=1$, we find that $w_1=0$ for the field-strength $F_{\mu_1\cdots\mu_{D/2}}$ of a $\tfrac{d}{2}$-form gauge field $A_{\mu_1\cdots\mu_{d/2}}$ in $D=d+2$ dimensions.
The conformal weight $w_s=\tfrac{d}{2}\,(s-1)$ under Weyl transformations is also in agreement with the scaling dimension $\Delta_s=s\,\left(\tfrac{d}{2}+1\right)-w_s=\tfrac{d}{2}+s$ under global conformal transformations (see e.g. Section.2.2 in \cite{Bekaert:2022ipg} for the formula relating the conformal weight $w$ and the scaling dimension $\Delta$, respectively for a conformal density and a conformal primary field). This is indeed the scaling dimension of a spin-$s$ singleton in $D$ dimensions, as will now be reviewed in the next subsection.
 
\subsection{Group-theoretical perspective}\label{groupth}

One of the most comprehensive work on the representation theory of spin-$s$ singletons in any dimension is  \cite{Angelopoulos:1997ij}. This work was the generalisation (to higher dimensions) of the seminal results from \cite{Angelopoulos:1980wg} on massless representations in four dimensions. Some of those results were reviewed in \cite[Section 3]{Bekaert:2011js} with notations and terminology which are more standard to high-energy physicists, and which we will follow here.

Singletons of integer spin $s\geqslant 1$ living in $D=d+2$ dimensions are those unitary irreducible representations of the Poincar\'e group $ISO_0(d+1,1)$ that lift (uniquely) to representations of the restricted conformal group $SO_0(d+2,2)$ \cite[Theorem 2.5]{Angelopoulos:1997ij}. Equivalently, singletons are those unitary irreducible representations of $SO_0(d+2,2)$ that remain irreducible upon restriction to a Poincar\'e subgroup $ISO_0(d+1,1)\subset SO_0(d+2,2)$.
It has been known for some time  \cite{Angelopoulos:1997ij,Siegel:1988gd} that spin-$s$ singletons identify with those lowest-weight representations ${\cal D}^{\mathfrak{so}(d+2,2)}(\frac{d}2+s\,;\,s,\ldots,s,\pm s)$ of the conformal algebra $\mathfrak{so}(d+2,2)$ that are induced from a conformal primary field of scaling dimension $\Delta_s=\frac{d}2+s$ and Lorentz structure labeled by a rectangular Young diagram made of $\frac{d}2+1$ rows of length $s$. Note that, for spin $s\geqslant 1$, singletons only exist in even dimension (i.e. for even $d$). These spinning singletons should correspond to the space of radiative solutions of generalised Bargmann-Wigner equations modulo subradiative ones. Indeed, the radiative boundary data at null infinity is a conformal Carrollian primary field carrying a unitary representation of the Poincar\'e group (see e.g. \cite{Donnay:2022wvx,West}).

Let us discuss in detail the scalar case ($s=0$), which exhibits many similarities but also some slight peculiarities. For instance, the representation ${\cal D}^{\mathfrak{so}(d+2,2)}(\frac{d}2,0)$ exists in any dimension. Upon restriction of the conformal algebra to a Poincar\'e subalgebra $\mathfrak{iso}(d+1,1)\subset\mathfrak{so}(d+2,2)$, this unitary irreducible module ${\cal D}^{\mathfrak{so}(d+2,2)}(\frac{d}2,0)$ of the conformal algebra reproduces the module $\mathcal{D}^{\mathfrak{iso}(d+1,1)}(\tfrac{d}{2},0)$ of the Poincar\'e algebra spanned by radiative modes of d'Alembert equation modulo subradiative ones. As explained in Section \ref{expnullinfty}, the latter representation is equivalent to the unitary irreducible representation of the Poincar\'e group $ISO_0(d+1,1)$  describing, \`a la Wigner, a massless scalar field in momentum representation. The representation ${\cal D}^{\mathfrak{so}(d+2,2)}(\frac{d}2,0)$ of the conformal algebra $\mathfrak{so}(d+2,2)$ lifts uniquely to a representation of the restricted conformal group $SO_0(d+2,2)$, since the latter group is connected and the representation is single-valued. As one can see, upon restriction of the restricted conformal group $SO_0(d+2,2)$ to a Poincar\'e subgroup $ISO_0(d+1,1)$, the representation ${\cal D}^{\mathfrak{so}(d+2,2)}(\frac{d}2,0)$ of $SO_0(d+2,2)$ becomes $\mathcal{D}^{\mathfrak{iso}(d+1,1)}(\tfrac{d}{2},0)$ of $ISO_0(d+1,1)$. Let us stress that this holds at the level of Lie group representations.

Now, let us consider instead the restriction of the restricted conformal group $SO_0(d+2,2)$ to a subgroup $SO_0(d+1,2)$, which can be interpreted as (the connected component of) the isometry group of the spacetime $AdS_{d+2}$. In general, the branching rule of the scalar singleton for this restriction is the following equality of $\mathfrak{so}(d+1,2)-$modules (see e.g. \cite[Proposition 3.1]{Angelopoulos:1997ij}):
\begin{equation}
\mathcal{D}^{\mathfrak{so}(d+2,2)}(\tfrac{d}{2},0)=\mathcal{V}^{\mathfrak{so}(d+1,2)}(\tfrac{d}{2},0)=\mathcal{D}^{\mathfrak{so}(d+1,2)}(\tfrac{d}{2},0)\inplus\mathcal{V}^{\mathfrak{so}(d+1,2)}(\tfrac{d}{2}+1,0) \,,    
\end{equation}
where $\mathcal{V}^{\mathfrak{so}(d+2,2)}(\tfrac{d}{2},0)$ denotes the space of radiative solutions to the Yamabe equation (which is the Klein-Gordon equation with critical mass-squared $m^2=-d(d+2)/(2R)^2$, where $R$ denotes the AdS curvature radius) and where the symbol $\inplus$ indicates that we are dealing with a semidirect\footnote{Note that the sum is direct in \cite[Proposition 3.1]{Angelopoulos:1997ij}. This appears to be due to the fact the maximal compact subalgebra $\mathfrak{so}(d+2)\oplus\mathfrak{so}(2)\subset\mathfrak{so}(d+2,2)$ is considered for the induction procedure in  \cite{Angelopoulos:1997ij}. Here, we prefer the usual  induction from the subalgebra $\mathfrak{so}(d+1,1)\oplus\mathfrak{so}(1,1)\subset\mathfrak{so}(d+2,2)$ for the sake of Lorentz-covariance. In fact, this choice is the one suitable for the flat limit, i.e. the In\"onu-Wigner contraction $\mathfrak{so}(d+1,2)\to\mathfrak{iso}(d+1,1)$.} sum of modules, i.e. the second summand is a submodule of the left-hand side, 
$\mathcal{V}^{\mathfrak{so}(d+1,3)}(\tfrac{d}{2}+1,0)\subset \mathcal{V}^{\mathfrak{so}(d+1,2)}(\tfrac{d}{2},0)$, while the first summand is the quotient $\mathcal{D}^{\mathfrak{so}(d+1,2)}(\tfrac{d}{2},0)=\mathcal{V}^{\mathfrak{so}(d+1,2)}(\tfrac{d}{2},0)/\mathcal{V}^{\mathfrak{so}(d+1,2)}(\tfrac{d}{2}+1,0)$.

There is an analogue of this branching rule for the Poincar\'e subgroup of the full conformal group $O(d+2,2)/\mathbb{Z}_2$. The latter contains, in particular, the inversion. As follows from Sections \ref{expnullinfty} and \ref{invduality}, the vector space $\mathcal{V}^{\mathfrak{iso}(d+1,1)}(\tfrac{d}{2},0)$ of all radiative solutions to d'Alembert equation carries an indecomposable representation of 
the Poincar\'e group $ISO_0(d+1,1)$ that lifts to an irreducible representation of the conformal group $O(d+2,2)/\mathbb{Z}_2$, which we will nonchalantly denote $\mathcal{D}^{\mathfrak{o}(d+2,2)}(\tfrac{d}{2},0)$. More precisely, the branching rule for this irreducible representation of the conformal group restricted to the usual Poincar\'e subgroup takes the form
\begin{equation}
\mathcal{D}^{\mathfrak{o}(d+2,2)}(\tfrac{d}{2},0)=\mathcal{V}^{\mathfrak{iso}(d+1,1)}(\tfrac{d}{2},0)=\mathcal{D}^{\mathfrak{iso}(d+1,1)}(\tfrac{d}{2},0)\inplus\mathcal{V}^{\mathfrak{iso}(d+1,1)}(\tfrac{d}{2}+1,0) \,, 
\end{equation}
where $\mathcal{V}^{\mathfrak{iso}(d+1,1)}(\tfrac{d}{2}+1,0)\subset\mathcal{V}^{\mathfrak{iso}(d+1,1)}(\tfrac{d}{2},0)$ is the $ISO_0(d+1,1)$-invariant subspace of subradiative solutions. This branching rule is in agreement with the detailed discussion of asymptotic solutions in Section \ref{expnullinfty}.

The inversion duality maps the usual Poincar\'e subgroup $ISO_0(d+1,1)\subset SO_0(d+2,2)$ to the exotic Poincar\'e subgroup $\widetilde{ISO}_0(d+1,1)\subset SO_0(d+2,2)$. Accordingly, the branching rule for the restriction of the conformal group to the exotic Poincar\'e subgroup takes the equivalent form
\begin{equation}
\mathcal{D}^{\mathfrak{o}(d+2,2)}(\tfrac{d}{2},0)=\mathcal{V}^{\widetilde{\mathfrak{iso}}(d+1,1)}(\tfrac{d}{2},0)=\mathcal{D}^{\widetilde{\mathfrak{iso}}(d+1,1)}(\tfrac{d}{2},0)\inplus\mathcal{V}^{\widetilde{\mathfrak{iso}}(d+1,1)}(\tfrac{d}{2}+1,0) \,,    
\end{equation}
where $\mathcal{V}^{\widetilde{\mathfrak{iso}}(d+1,1)}(\tfrac{d}{2}+1,0)\subset \mathcal{V}^{\widetilde{\mathfrak{iso}}(d+1,1)}(\tfrac{d}{2},0)$ is the $\widetilde{ISO}_0(d+1,1)$-invariant subspace of radiative solutions that are evanescent at the null cone. This branching rule is in agreement with the detailed discussion of solutions near the null cone in Section \ref{expnullcone} and with the comments at the end of Section \ref{invduality} on the inversion duality.

\section{Conclusion}\label{concl}

In this paper, we argued that if one takes the full global conformal symmetry seriously, then the subradiative solutions of wave equations play a role that is equally important to the radiative ones. In particular, the inversion relates these two branches of solutions. Although interactions and curvature generically break\footnote{Note that there exists some remarkable exceptions. For instance, extremal back holes admit the Couch-Torrence inversion \cite{CouchTorrence} exchanging the horizon of an extremal black hole with null infinity. All our qualitative observations will extend to wave equations with Weyl symmetry (such as Yamabe and Maxwell equations) in such a background metric (see e.g. \cite{Bizon:2012we}). For instance, the boundary data of the subradiative sector should be a Carrollian field on the horizon, which is exchanged, via a Couch-Torrence inversion, with the boundary data at null infinity of the radiative sector.} the inversion symmetry, it is tantalising to postulate that the main qualitative features of the subradiative sector will survive. For instance, that the boundary data of the tower of subradiative modes is a Carrollian field defined on a null cone $\mathscr{N}_{d+1}^+$ in the infinitesimal vicinity of null infinity $\mathscr{I}_{d+1}^+$, emitted from a cut $\mathscr{N}_{d+1}^+\cap\mathscr{I}_{d+1}^+\cong S^d$. In spacetime dimension four, this is in agreement with the literature (see e.g. Theorem 18.1 and Figure 18.2 in \cite{Kroon:2016ink}). Furthermore, the inversion symmetry of massless spin-two equations in four spacetime dimension suggest that, at least for linearised gravity, the subradiative modes can be described geometrically as an (equivalence class) of Carrollian connections  on $\mathscr{N}_3^+$ (similarly to the description of radiatives modes on $\mathscr{I}_3^+$, cf. \cite{Ashtekar:1987tt}). 
In fact, it would be interesting to explore the idea that some generalisation of the asymptotic quantisation program, extended to the subradiative sector, might provide some new insights in the Carrollian approach to flat holography.

We have also shown that the asymptotic analysis of solutions of wave equations with Weyl symmetry (such as massless fields of any spin in four dimension) simplifies dramatically in flat Bondi coordinates on Minkowski spacetime,
because a Weyl transformation allows to map the original asymptotic characteristic problem (i.e. the holographic description of bulk solutions in terms of boundary data at null infinity and on an interior null cone) to a much simpler problem, which is a Goursat problem (where the boundary data is fixed on two intersecting null hyperplanes). This suggests that a Hamiltonian approach to the asymptotic analysis should admit an elegant reformulation in terms of the recent Hamiltonian approach to Dirac's lightfront formulation \cite{Barnich:2024aln,Gonzalez:2023yrz} where the constraints and subtleties related to zero modes are carefully addressed and would correspond to subradiative modes.
This reformulation should in principle extend to the quantum level, at least for free theories in a flat background, possibly providing a fruitful bridge between current infrared investigations and the characteristic initial value problem in lightfront quantisation (cf. the motivations in \cite{Barnich:2024aln}).

We have obtained an explicit reconstruction formula relating the boundary data with the bulk solution. It would be nice to find a direct relation between this boundary-to-bulk formula on Minkowski spacetime to its celebrated analogue in AdS spacetime, via a suitable flat limit. One should note that the Huygens principle, for even spacetime dimension, is not manifest in our formula (which does not make any distinction between even and odd $d$). In fact, an interesting direction to explore would be to find the precise relation between our formula and the celebrated Kirchhoff--d'Adh\'emar formula (see \cite{Penrose:1980yx} for the generalisation of the latter to any spin) which is an integral over a sphere only (basically, over the cut $\mathscr{N}_3^+\cap\mathscr{I}_3^+\cong S^2$).

Finally, there are several topics of general interest, that we have not touched in our analysis and that would deserve to be studied in the future. An incomplete list of things that we left for future investigations are:
\begin{itemize}
    \item the analogue of the antipodal matching condition (cf. \cite{Strominger:2017zoo,Strominger:2013jfa} and refs therein) in the subradiative sector,
    \item the (ir)regularity structure of the subradiative sector at spatial and null infinity,
    \item the expected relation between these last two points, in particular in the Hamiltonian formalism (see e.g. the paper \cite{Henneaux:2019yax} reviewing the series of work of these authors on this issue),
    \item the limit where the vertex of the interior null cone (on which the subradiative data is prescribed) goes to timelike infinity, i.e. either $\mathscr{N}_{d+1}^+\to\mathscr{I}_{d+1}^-$ or $i^+$.\footnote{The first limit ($\mathscr{N}_{d+1}^+\to\mathscr{I}_{d+1}^-$) is of interest for scattering. Note that, in the regular case, the asymptotic characteristic problem goes (in the second limit, $\mathscr{N}_{d+1}^+\to i^+$) to a characteristic problem on a cone (see e.g. \cite{Kroon:2016ink}). It might be interesting to investigate what the subradiative sector adds to this picture. The interplay between the second and fourth points in the above list would also be of interest.}
\end{itemize}

\section*{Acknowledgments}

We would like to thank particularly Yannick Herfray for illuminating comments and exchanges. We also thank Glenn Barnich, Andrea Campoleoni, Ma\"el Chantreau, Laura Donnay, Yegor Goncharov, Marc Henneaux, Sucheta Majumdar, Simon Pekar and Jean-Philippe Nicolas for useful discussions.

\appendix

\section{Two characteristic initial value problems}\label{regularity}

Let us stress some similarities and important differences between the mathematical problem \eqref{subspaceasympt2} considered in Section \ref{expnullcone}, the so-called ``asymptotic characteristic problem'' for d'Alembert equation, and another standard boundary problem for d'Alembert equation, the so-called ``characteristic problem on a cone''. These two boundary problems are both examples of ``characteristic initial value problem'' but it is crucial to distinguish them (see e.g. the textbook \cite{Kroon:2016ink}, in particular Figure 18.1).

It is standard to look for solutions $\upphi$ of d'Alembert equation $\Box_{{}_{\mathbb{R}^{d+1,1}}}\upphi =0$ that are smooth in the causal future $J_{d+2}^+(0)$ of the origin and whose restriction to the future null cone $\mathcal{N}^+_{d+1}=\partial J_{d+2}^+(0)$ is a prescribed function $\phi$ on $\mathcal{N}^+_{d+1}$ which is smooth down to its vertex.\footnote{For $d=0$, this characteristic problem on a cone coincides with the standard Goursat problem \cite{Courant} for the wave equation (in $D=2$ dimensions). For this reason, the characteristic problem on a cone is sometimes called ``Goursat problem'' among the mathematical relativity community (see e.g. \cite{Mason:2004lqj} where the null cone is past null infinity $\mathcal{I}^-_{d+1}$).} 
This \textit{characteristic problem on a cone} is well posed: there exists a unique solution $\upphi$ for each given $\phi$ (see e.g. \cite[Theorem 5]{Rendall} for this classical result). However, we saw that two independent functions have to be prescribed for the asymptotic characteristic problem to be well posed, cf. \eqref{subspaceasympt2}. 

There is no contradiction between these two results since they correspond to distinct characteristic initial value problems. For $d\geqslant 1$, the behaviour in \eqref{subspaceasympt2} shows that the restriction $\phi(r,\mathbf y)=\upphi(u=0,r,\mathbf y)=\frac1{r^{\frac{d}{2}}}\varphi_0(r,\mathbf y)$ to the null cone is, in general, \textit{not} a smooth function on $\mathscr{N}_{d+1}$. More precisely, the function $\phi$ is clearly singular on the null line $r=0$ (i.e. the line $x^\mu=u\,n^\mu$) inside the null cone for a generic function $\varphi_0$.\footnote{Remember from Section \ref{flatBondicords} that $r=0$ is the equation of the null line $x^\mu = u\, n^\mu$, which is the intersection between the hyperplane $x^+=0$ and the null cone $x^2=0$.}
Of course, this function $\phi$ can be (trivially) smooth if it vanishes identically ($\phi\equiv 0$), i.e. when the solution is evanescent ($\varphi_0\equiv 0$).

There exists non-trivial radiation fields which are evanescent but the price to pay is that they will not be regular everywhere. More precisely, such solutions can not be smooth everywhere in the interior of Minkowski spacetime, or they would have to vanish identically (by the uniqueness theorem of smooth solutions). Therefore, radiation fields which are evanescent on the null cone will become singular somewhere in the interior of Minkowski spacetime, e.g. on the null hyperplane $x^+=0$ (where sits the coordinate singularity of flat Bondi coordinates).

In any case, the present paper does not even attempt to address seriously the difficult issues of existence and regularity of solutions. Our ansatze \eqref{asexp1/r} and \eqref{expu} should merely be taken as formal power series. In coordinate-independent terms, they are nothing but the jets of infinite order of solutions $\upphi$ in the transverse directions to the good cut. In other words,  one may say, to be mathematically precise (if not pedantic), that our investigations in this paper are restricted to the study of formal solutions to d'Alembert equation in the infinitesimal neighborhood\footnote{See e.g. \cite[p.6]{Deligne} for a definition of this notion.} (of infinite order) of the cut $\mathscr{N}_{d+1}^+\cap\mathscr{I}_{d+1}^+\cong S^d$ on the boundary of the conformal compactification of Minkowski spacetime $\mathbb{R}^{d+1,1}$. 

\section{Proof of redundancy
}\label{Maxapp}

We have to show here that equations \eqref{max_5} and \eqref{max_6} are redundant when $n \neq 0$ and when $n \neq d - 2$, respectively. 

Taking the time derivative of \eqref{max_1}, one gets
\begin{equation}\label{using}
    n\,\dot{F}_{ij}^{(n)} + \partial{}^{}_{[i} \dot{F}_{j]v}^{(n-1)}  = 0   
\end{equation}
and using $\eqref{max_3}$, leads to
\begin{equation} \partial{}^{}_{[i}\dot{F}_{ j] v}^{(n-1)}  - n\, \partial{}^{}_{[i}F_{j]u}^{(n)} = 0  \stackrel{\eqref{using}}{\implies} n \left(\dot{F}_{ij}^{(n)} + \partial{}^{}_{[i}F_{j]u}^{(n)}  \right)  = 0\label{redund1}\end{equation}
which is equivalent to \eqref{max_5} when $n\neq 0$.
Now, taking the time derivative of $\eqref{max_2}$ gives
\begin{equation}\label{b3}
    (2 - d + n) \dot{F}_{vu}^{(n)} = \frac{1}{2} \partial^i \dot{F}_{iv}^{(n-1)}
\end{equation}
And taking the divergence of \eqref{max_4} leads to
\begin{equation}
     (2 - d + n)\partial^i F_{ui}^{(n)} + \partial^i\dot{F}_{vi}^{(n-1)} + \frac{1}{2} \partial^i\partial^j F_{ji}^{(n-1)} = 0  \stackrel{\eqref{b3}}{\implies} (2 - d + n)\left( \dot{F}_{vu}^{(n)} + \frac{1}{2} \partial^i F_{iu}^{(n)} \right) = 0 \label{redund2}
\end{equation}
which is equivalent to \eqref{max_6} when $2 - d + n\neq 0$.
So we see that \eqref{redund1} and \eqref{redund2} prove our claim.

\section{Boundary-to-bulk propagator}\label{Kprop}

To compute the integral \eqref{K(uz)}, one can expand the second exponential in powers of $1/\omega$:
\begin{eqnarray}
K(y) &=&  \frac{1}{2\pi}\int d\omega \,e^{-i\omega y}\, e^{+\frac{i}{\omega}} =\frac{1}{2\pi} \int d\omega\, e^{-i\omega y}  + \frac{1}{2\pi}\sum_{n=1}^{\infty} \frac{1}{n!}\int d\omega\, e^{-i\omega y} \,\left(\frac{i}{\omega}\right)^n    \\
&=&  \delta(y)  + \frac{1}{2\pi}\sum_{n=1}^{\infty} \frac{i^n}{n!}\int d\omega\, e^{-i\omega y}\,\frac{(-1)^{n-1}}{(n-1)!} \frac{d^{n-1}}{d{\omega'}^{n-1}}\left(\frac{1}{\omega'}\right)\\
&=&  \delta(y)  + \frac{1}{2\pi}\sum_{n=1}^{\infty} \frac{i^n}{n!}\int d\omega\, e^{-i\omega' y}\,\frac{(-iy)^{n-1}}{(n-1)!} \left(\frac{1}{\omega'}\right)\\
&=& \delta(y) + \frac{1}{2}\sum_{n=1}^{\infty} \frac{1}{n!}\frac{y^{n-1}}{(n-1)!} sgn(y)
\end{eqnarray}
For more details, look at \cite[Chapter 6.4, Example 3]{Kanwal}. We can express the above sum in terms of the well known Bessel function of the first kind
\begin{equation}
J_0(2i\sqrt{y})  = \sum_{n=0}^{\infty} \frac{y^{n}}{(n!)^2} \implies \frac{d}{dy}\Big(J_0(2i\sqrt{y})\Big) = \sum_{n=1}^{\infty} \frac{y^{n-1}}{n!(n-1)!}
\end{equation}
This yields,
\begin{equation}
K(y) =  \delta(y) + \frac{1}{2} sgn(y) \frac{d}{dy}\Big(J_0(2i\sqrt{y})\Big)\,.    
\end{equation}

\section{Complete reconstruction formula}\label{heuristic}

The goal of this appendix is to generalise the results in Section \ref{reconstructionform} to the subradiative modes.
The main idea is to follow the leitmotiv of this paper by addressing the equivalent (but simpler) higher-dimensional Goursat problem \eqref{Goursatlightcone} for the rescaled field $\tilde\upphi$ in light-cone coordinates $(u,v,\mathbf y)$. 

One can start by rewriting the compact form \eqref{explicitrad} of the radiation field in terms of the rescaled field:
\begin{equation}\label{radexp}
    \tilde\upphi_{\text{rad}}(u,v,\mathbf y) = e^{-\,v\,\frac{\partial_i\partial^i}{4}(\partial_u)^{-1}}\,\phi_0(u, \mathbf y) \,.
\end{equation}
Applying the inversion duality $u\leftrightarrow v$ suggests that the subradiative piece in the rescaled general solution reads
\begin{equation}
    \tilde\upphi_{\text{sub}}(u,v,\mathbf y) = e^{-\,u\,\frac{\partial_i\partial^i}{4}(\partial_v)^{-1}}\,\psi_0(v, \mathbf y) \,,
\end{equation}
which agrees with \eqref{subexp}-\eqref{tail'} if one takes the prescription 
\begin{equation}
(\partial_v)^{-m}(v^n)=\frac{n!}{(m+n)!}\,v^{m+n}\,.
\end{equation}

Another way to generalise the results of Section \ref{reconstructionform} is to consider  the wave equation in light-cone-like coordinates, i.e. \eqref{dAlembt}.
We look for separable solutions of the form $\tilde\upphi(u,v,\mathbf y) = U(u)\,V(v,\mathbf y) $,
\begin{equation}
\partial_v V (v,\mathbf y)U'(u) = -\frac{1}{4} \partial_i\partial^i V(v,\mathbf y) U(u) \implies  \frac{U'(u)}{U(u)} = -\frac{\partial_i\partial^i V(v,\mathbf y)}{4 \partial_v V(v,\mathbf y)}  = -i\omega
\end{equation}
as they are both independent functions of $u$ and $v,\mathbf y$ respectively. 
Thence, the equations are
\begin{equation}
    U'(u) = -i\omega U(u)\,, \qquad \partial_v V(v,\mathbf y) = \frac{-i\partial_i\partial^i}{4\omega} V(v,\mathbf y)\,.
\end{equation}
We can now solve for $U(u)$ and $V(v,\mathbf y)$, and get the separable solution $\tilde\upphi(u,v,\mathbf y) =U(u) V(v,\mathbf y) = \exp[-iu\omega -iv\frac{\partial_i\partial^i}{4\omega}] f(\omega,\mathbf y)$. Integrating over the parameter, yields
\begin{equation}
\tilde\upphi_{\text{rad}}(u,v,\mathbf y)  = \int_{-\infty}^\infty d\omega\, e^{-i\left(u\,\omega\, +\,v\,\frac{\partial_i\partial^i}{4 \omega}\right)} f(\omega,\mathbf y)    \,,
\end{equation}
which reproduces our reconstruction formula \eqref{beautiful}, by identifying $f(\omega,\mathbf y) = \omega^{\frac{d}{2} - 1} \phi(\omega,\mathbf y) $.
But we could also have looked for separable solutions of the form $\tilde\upphi(u,v,\mathbf y) =V(v)U(u,\mathbf y)$. This would have given $\tilde\upphi(u,v,\mathbf y) 
= 
 \exp[-iv\omega -iu\frac{\partial_i\partial^i }{4 \omega}] g(\omega,\mathbf y)$.
Adding both option gives us,
\begin{equation}\label{fullreconstr}
\tilde\upphi(u,v,\mathbf y)  = \int_{-\infty}^\infty d\omega\, \left[e^{-i\left(u\,\omega\, +\,v\,\frac{\partial_i\partial^i}{4 \omega}\right)} f(\omega,\mathbf y)  +e^{-i\left(v\,\omega\, +\,u\,\frac{\partial_i\partial^i}{4 \omega}\right)} g(\omega,\mathbf y)    \right]\,.
\end{equation}
In this way, the general solution indeed depends of two independent functions $f(\omega,\mathbf y)$ and $g(\omega,\mathbf y)$. The boundary data of the Goursat problem \eqref{Goursatlightcone} is related to these two functions as follows:
\begin{eqnarray}
\phi_0(u,\mathbf y) & = &\int_{-\infty}^\infty d\omega\, \left[e^{-i\,u\,\omega} f(\omega,\mathbf y)  +e^{-i\,u\,\frac{\partial_i\partial^i}{4 \omega}} g(\omega,\mathbf y)    \right]\,,\\
\varphi_0(v,\mathbf y)  &=& \int_{-\infty}^\infty d\omega\, \left[e^{-i\,v\,\frac{\partial_i\partial^i}{4 \omega}} f(\omega,\mathbf y)   +e^{-i\,v\,\omega} g(\omega,\mathbf y) \right]\,.
\end{eqnarray}
In this way, one may also try to recover the above formulae \eqref{radexp}-\eqref{subexp}.

Another argument in favor of the complete reconstruction formula \eqref{fullreconstr} is as follows. In light-cone coordinates, one gets
\begin{equation}
\phi(x) = \int dp^+ dp^- d^d\mathbf{p} \,\delta\left(-2p^+p^- + p_ip^i\right)e^{-ip^+x^- - ip^-x^+ + ip_ix^i }\phi(p)
\end{equation}
The next step is to integrate with respect to $p^-$ (or $p^+$) by using the heuristic identity\footnote{This formula should be understood in the sense of distributions, as a particular instance of the formula:
\begin{equation}
\int_{\mathbb{R}^n} f(\mathbf{x})\, \delta\big(g(\mathbf{x})\big) \,d^n\mathbf{x}=\int_{g^{-1}(0)} \frac{f(\mathbf{x})}{|\mathbf{\nabla} g|}\, d\sigma(\mathbf{x})
\end{equation}
where $f$ is a test function while $g$ is a real function on $\mathbb{R}^n$ and $d\sigma$ is the measure on the preimage $g^{-1}(0)\subset \mathbb{R}^n$ made of all zeros of the function $g$.}
\begin{equation}
\delta\left(-2p^+p^- + p_ip^i\right)=\frac{1}{2\abs{p^+}}\,\delta\left(p^- - \frac{p_ip^i}{2p^+}\right)
+\frac{1}{2\abs{p^-}}\,\delta\left(p^+ - \frac{p_ip^i}{2p^-}\right)\,.    
\end{equation}
This leads to
\begin{eqnarray}
\label{delta_prob4}\phi(x) &=& \int dp^+  d^d\mathbf{p}\,  \frac{1}{2\abs{p^+}} \,e^{-ip^+x^- - i\frac{p_ip^i}{2p^+}\,x^+ + ip_ix^i }\phi\left(p^+,p^- =\frac{p_ip^i}{2p^+},p^i \right)\nonumber\\
&&+\int dp^-  d^d\mathbf{p}  \,\frac{1}{2\abs{p^-}}\,e^{-ip^-x^+ - i\frac{p_ip^i}{2p^-}\,x^- + ip_ix^i }\phi\left(p^+ =\frac{p_ip^i}{2p^-},p^-, p^i \right)\,. \label{sumtwoint}
\end{eqnarray}
Setting $p^+=p^-=\omega$ in \eqref{sumtwoint}, one obtains the reconstruction formula by expressing the right-hand-side in terms of the Fourier transform over the celestial plane.

\pagebreak


\end{document}